\documentstyle[12pt,aaspp4]{article}

     \newcommand{\csxfv}{[1965\AA]}
     \newcommand{\csxoh}{[1960\AA]}

\begin{document}

     \title {Convection, Thermal Bifurcation, and the Colors of A
     Stars}

     \author {Theodore Simon\altaffilmark{1}}
     \affil{Institute for Astronomy, University of Hawaii, 2680
     Woodlawn Drive, Honolulu, HI 96822}

     \and

     \author{Wayne B. Landsman\altaffilmark{1}}
     \affil{Hughes STX Corp., NASA/GSFC, Code 681, Greenbelt, MD
     20771}

     \altaffiltext{1}{Guest Observer with the {\it International
     Ultraviolet Explorer (IUE)} satellite}

     \begin{abstract}

     Broad-band ultraviolet photometry from the {\it TD-1}
     satellite and low dispersion spectra from the short
     wavelength camera of {\it IUE} have been used to investigate a 
     long-standing proposal of B\"ohm-Vitense that the normal main
     sequence A-- and early--F stars may divide into two different
     temperature sequences:  (1) a high temperature branch (and
     plateau) comprised of slowly rotating convective stars, and (2) a low
     temperature branch populated by rapidly rotating radiative stars.
     We find no evidence from either dataset to support such a claim, or to
     confirm the existence of an ``A--star gap'' in the \bv\ color range 
     $0.22 \leq \bv \leq 0.28$ due to the sudden onset of convection.

     We do observe, nonetheless, a large scatter in the 1800--2000 \AA\
     colors of the A--F stars, which amounts to $\sim0.65$ mags at a
     given \bv\ color index.  The scatter is not caused by interstellar
     or circumstellar reddening.  A convincing case can also be made
     against binarity and intrinsic  variability due to pulsations of
     $\delta$ Sct origin.  We find no correlation with established
     chromospheric and coronal proxies of convection, and thus no
     demonstrable link to the possible onset of convection among the 
     A--F stars.  The scatter is not instrumental.  Approximately 0.4
     mags of the scatter is shown to arise from individual differences 
     in surface gravity as well as a moderate spread (factor of $\sim3$)
     in heavy metal abundance and UV line blanketing.  A dispersion of  
     $\sim0.25$ mags remains, which has no clear and obvious explanation.  
     The most likely cause, we believe, is a residual imprecision in our
     correction for the spread in metal abundances.  However, the
     existing data do not rule out possible contributions from
     intrinsic stellar variability or from differential UV line 
     blanketing effects owing to a dispersion in microturbulent velocity.
     \end {abstract}

     \section{Introduction}

     Along the lower half of the main sequence, it is among the A-type
     stars that a shallow convection zone first appears in the outer
     envelope of low mass stars (e.g., Gilliland 1986). For a variety
     of observational and astrophysical reasons, it has proved quite
     difficult to find the precise location where, in terms of effective
     temperature, \bv\/ color, or spectral class, this structural change
     actually occurs. The primary spectral proxies for convection, namely,
     X-ray and near ultraviolet (UV) emission from hot coronal and
     chromospheric gas, suggest that convection is common, though perhaps
     not universal, among the late A--type and early F--type stars
     (hereinafter, A--F stars), but is much more infrequent among the
     middle--A stars (Simon \& Landsman 1991; Simon, Drake, \& Kim 1995).
     Why it is that some A or F stars are X-ray or UV emitters, and are
     thus demonstrably convective, while others are not, remains a mystery.

     This perplexing issue was addressed in a series of articles begun
     nearly three decades ago by B\"ohm-Vitense (1982, hereinafter
     BV82, and references cited therein).  A principal conclusion of those
     studies is that when main sequence convection first sets in among
     the A--F stars, it does so very abruptly. As a result of this
     change, according to B\"ohm-Vitense, there may be major effects on
     the spectral line profiles of a star (B\"ohm-Vitense 1970a), as well
     as pronounced changes in the overall spectral energy distribution.
     At visible wavelengths, within the Paschen continuum below 5000\AA,
     for instance, the onset of convection makes a star look redder, and
     thus cooler, than it would otherwise appear in the absence of
     convection (B\"ohm-Vitense 1970b, 1978).  And because the
     transformation is expected to take place over just a narrow range in
     mass, there is a large increase in the \bv\/ color for only a
     slight drop in the effective temperature, $T_{\rm eff}$.
     Consequently, in a plot of the observational main sequence as a
     function of \bv, a plateau in $T_{\rm eff}$ should appear in the
     midst of the A--F star range, extending from \bv = 0.22 to \bv =
     0.28 (B\"ohm-Vitense \& Canterna 1974; B\"ohm-Vitense 1981a,
     BV82).

     The way in which these effects of convection alter A--F star
     spectra is either by  modifying the temperature gradient in deep
     layers of the atmosphere (B\"ohm-Vitense 1978), or else through
     the mediation of photospheric granulation (BV82).  Support for the
     latter explanation, cited in BV82, comes from numerical simulations 
     by Nelson (1980), which suggest that the temperature and pressure 
     inhomogeneities associated with a solar-type granulation pattern 
     can redden the \bv\/ color of an A or early F star by as much as
     0.07 mag.  Modern studies of convection based on the analysis and
     modelling of spectral line bisectors all seem to imply, however, 
     that the granulation patterns of these early stars and the Sun may bear
     very little resemblance to one another (e.g., Gray 1989, Dravins 1990).

     Other work by B\"ohm-Vitense \& Canterna (1974) and BV82 goes on
     to suggest that if the incipient convection is weak, it may be
     suppressed by rapid rotation.  Hence, rapidly rotating A--F
     stars should tend to stay radiative, exhibiting normal spectra 
     and colors, while slowly rotating stars, offering no
     obstacle to convection, should tend to become convective and be
     shifted to redder \bv\/ colors.  This differentiation based on
     rotation should therefore give rise to two distinct sequences of 
     A--F stars in the H--R plane:  (1) a high--$T_{\rm eff}$ branch
     populated by the slowly rotating, weakly convective stars, and,
     at the same \bv\/ colors, (2) a low--$T_{\rm eff}$ branch comprised
     of the rapidly rotating, purely radiative stars.  Farther down
     the main sequence, once convection is much more firmly established
     among the middle and late F stars at \bv $\geq$ 0.30, these two
     branches merge back together again to form a single temperature
     sequence.

     The existence of a thermal bifurcation among the A--F stars
     provides, according to B\"ohm-Vitense (1970b) and B\"ohm-Vitense
     \& Canterna (1974), a natural explanation for the deficiency some
     authors find in the numbers of late A stars relative to those of
     earlier and later spectral classes, both in the counts of stars
     in the solar neighborhood and also in cluster stars of various
     ages (e.g., Houk \& Fesen 1978, Harris et al. 1993).  In their
     view, it is the sudden onset of convection near \bv = 0.22, and
     the resulting shift in \bv\/ color, that creates this apparent
     ``A--star gap''.  However, other observers (e.g., Kjeldsen \&
     Frandsen 1991) report they are unable to confirm that such a gap 
     truly exists.

     A test of these ideas concerning the onset of convection and its
     observable effects on stellar energy distributions was presented
     in BV82.  The analysis carried out there was based on the premise
     that (1) the main spectroscopic changes induced by convection are
     limited to visible wavelengths, and that (2) none of the effects of
     convective granulation predicted by Nelson's (1980) calculation
     can penetrate into the layers of the photosphere from
     which the UV spectrum of an A--F star emerges.  Thus, the UV
     energy distribution of an A--F star, unlike its optical spectrum,
     should reflect the true $T_{\rm eff}$ of the star. In particular, 
     from a consideration of theoretical model atmosphere fluxes, it
     was shown that the ratio of the apparent brightness of a star at
     1900\AA\ to its brightness at optical wavelengths (or the color
     index formed therefrom) provides a sensitive estimate of
     $T_{\rm eff}$, which is free of the effects that might arise
     with the onset of convection.

     Narrow-band ($\sim40$\AA\ resolution) fluxes measured at 1900\AA\
     by the S2/68 Sky Survey Telescope aboard the
      {\it TD--1} satellite (Jamar et al. 1976, Macau-Hercot et
     al. 1978) were used in BV82 to derive effective temperatures for
     a sample of A and F stars by comparison with a grid of model
     atmospheres.  The resulting $T_{\rm eff}$ estimates were then
     plotted against the observed \bv\/ colors.  In the diagram thus 
     derived (see Figure 7 of BV82), B\"ohm-Vitense identified two
     branches of stars as we have just described them:  an upper one
     comprised of slowly rotating, convective stars that have been
     shifted horizontally to the right in \bv\/ color to form a
     temperature ``plateau,'' and a lower one occupied by rapidly
     rotating stars that have remained fully radiative.  Our version
     of this diagram, which omits the high luminosity giants and
     bright giants included in the original plot, is shown here as
     Figure 1 (see also Saxner \& Hammarb\"ack 1985). 

     Even at its widest separation, the vertical distance between the
     two temperature sequences is no greater than $\Delta T_{\rm eff}
     = 500$ K, which corresponds to a spread in UV flux of less than a 
     factor of two.  By comparison, the photometric errors in the
     narrow-band {\it TD--1} fluxes used in BV82 to construct this diagram 
     range from 5\% to more than 40\%; fully one-third of the errors,
     as cataloged by Jamar et al. (1976) and Macau-Hercot et al. (1978),
     are above $\pm25$\%.  Errors this large imply an uncertainty of at
     least $\pm180$ K in the $T_{\rm eff}$ estimates of individual stars.
     Given uncertainties of this size, the reality of the proposed
     temperature plateau and temperature bifurcation was not, in our
     opinion, convincingly proved.

     There are two additional reasons for caution that deserve mention
     here. First, for the same stars analyzed in BV82 by the UV method,
     independent estimates of $T_{\rm eff}$ can be derived from
     $uvby\beta$ photometry (Moon \& Dworetsky 1985; Napiwotzki,
     Sch\"onberner, \& Wenske 1993; Smalley \& Dworetsky 1995).  In a
     number of cases, this leads to a substantial disagreement in the
     $T_{\rm eff}$ estimates.  As illustrated in Figure 1, some stars on 
     the upper convective branch in BV82  belong on the lower radiative 
     branch according to their $uvby\beta$ photometry, while others
     should move in the opposite direction.  These inconsistencies are
     substantial enough, we think, to raise doubt about whether the
     temperature gap suggested by the narrow-band {\it TD--1} photometry
     is entirely credible.  Whether this apparent contradiction is actually
     significant depends, however, on the accuracy of the individual
     temperature estimates.
     Based on their ability to recover the $T_{\rm eff}$ of the fundamental
     stars used to calibrate their technique, Moon \& Dworetsky (1985)
     estimate the external accuracy of their method to be no better than 
     $\pm250$ K, while similar evaluations have been cited by Smalley
     \& Dworetsky (1993) and Napiwotzki et al. (1993).  Such a large
     uncertainty significantly reduces the confidence with which a
     temperature differential amounting to $\Delta T_{\rm eff} = 500$ K
     might be detected by means of the $uvby\beta$ photometry.
 
     In addition, we note that while the effects of convection on 
     stellar energy distributions and the resulting spread in photometric
     colors are presumed to be limited to the A--F stars, the issue of 
     whether hotter or cooler stars show as great a spread in their 
     photometry has not been systematically considered before.  Figure 2
     is a plot of the narrow-band 1900\AA\ colors for the main sequence
     stars listed in Tables 1 and 2 of BV82.  Also plotted are the
     colors predicted by ATLAS9 model atmospheres from Kurucz (1993a)
     for solar metallicity and $\log g = 4$, 
     and offset parallel lines which demarcate a band $\pm0.4$ mags
     wide above and below on either side of the model sequence.  As
     mentioned earlier, and as explained more fully in BV82, any
     vertical spread in the UV color index is equivalent to a spread
     in $T_{\rm eff}$.  A considerable scatter is evident along the
     main sequence in the UV emission of stars on either side of the
     ``A-star gap,'' both above it and below it.  This is perhaps not
     surprising for the cooler stars at \bv $>0.3$, given the increasing
     faintness of these stars in the UV and the greater complexity of their
     spectra due to line blanketing.  On the other hand, it is harder to
     reconcile the dispersion shown by the hotter stars, considering that
     the scatter extends all the way up the main sequence to the earliest
     points plotted near \bv = 0.0, among stars for which there is no
     suspicion that convection is present.  Scatter
     of similar amplitude, equivalent to 500 K in $T_{\rm eff}$, had
     already been noted before at \bv $<0.1$ in narrow-band {\it TD--1}
     photometry at 1400\AA\ and 2100\AA\ by Malaise et al. (1974),
     who attributed it to individual variations in spectral line
     blanketing, and also in Am-type stars by van 't Veer-Menneret et
     al. (1980).  Clearly, then, a number of causes other than 
     convection (including observational errors) may need to be
     considered as possible explanations for the observed spread in
     the {\it TD--1} fluxes of the A- and F-type stars.

     Given the apparent conflicts in  stellar temperature estimates
     made independently from UV and optical photometry, and the
     scatter in UV flux present even among the fully radiative stars
     of Figure 2, we re-examine in this paper the related issues of
     the onset of convection and the effective temperature scale of
     the A--F stars --- two important questions that were raised by 
     B\"ohm-Vitense's pioneering studies.  Initially we discuss our 
     analysis of an enlarged sample of {\it TD--1} flux measurements for 
     these stars, in which we make use of broad-band ($\sim330$\AA)
     fluxes from {\it TD--1}.  These data are generally much more
     accurate than the narrow-band fluxes used in BV82.  We then go on
     to discuss narrow-band (20--40\AA) spectrophotometry from the
     {\it IUE} telescope, which we obtained for a sample of A--F stars
     in the 1800--2000\AA\ region.  In this work we take advantage of
     the high sensitivity and photometric accuracy of {\it IUE}.  All
     of the A or F stars that were observed only with great difficulty by
     {\it TD--1} are within the grasp of {\it IUE} in exposure times
     of just a few minutes, or even a few seconds.  Moreover, the
     photometric uncertainties of the {\it IUE} data, both internal
     and external, are at a level of a few percent, corresponding
     to an accuracy of $\pm25$ K in $T_{\rm eff}$.  This provides a
     vast improvement over the narrow-band {\it TD--1} spectrophotometry
     upon which B\"ohm-Vitense relied.

     \section{Spectrophotometry from {\it TD--1}}

     Using an electronic version of the Thompson et al. (1978)
     catalog, we extracted broad-band ($\Delta \lambda \approx
     330$\AA) {\it TD--1} fluxes centered at 1965\AA\ for more than 
     600 stars in the {\it Yale Bright Star Catalogue} (Hoffleit \&
     Jaschek 1982) with \bv\/ colors in the range $0.00 \leq \bv \leq 0.45$.
     From this initial list we eliminated all known Ap and Am stars,
     and any normal star with a cataloged flux error $>10$\%.  For
     each of the remaining $300+$ stars we obtained Str\"omgren
     ($uvby\beta$) photometry either from the Hauck \& Mermilliod
     (1990) catalog or from the SIMBAD database.  We used this
     photometry to estimate $T_{\rm eff}$ in accordance with the
     method developed by Moon \& Dworetsky (1985), which, for stars
     cooler than 8500 K, relies mainly on the H$\beta$ index, $\beta$, 
     to gauge the value of $T_{\rm eff}$.  A plot of the temperatures
     determined in this way, restricted to those stars in our {\it TD--1}
     sample with colors in the range $0.15 \leq \bv \leq 0.33$, is shown
     in Figure 3.  No obvious clustering of the stars into two separate 
     temperature sequences is apparent here.  However, considering the
     acknowledged uncertainty of $\pm250$ K in our temperature estimates,
     a gap of the size claimed by BV82 might well be hidden within the
     cloud of points plotted in Figure 3.  

     A more fundamental issue is whether one should expect to see 
     evidence of a temperature bifurcation in a plot such as Figure 3, 
     where both axes are potentially affected by convection. The profile  
     of H$\beta$ is known to be sensitive to the influence of convection 
     (see, e.g., van 't Veer-Menneret \& M\'egessier 1996).  According 
     to B\"ohm-Vitense (1970a), ``\ldots stratification makes a convective
     star appear redder and the hydrogen lines somewhat weaker than in a
     radiative star of the same $T_{\rm eff}$. This still holds when
     temperature inhomogeneities in the convective star are considered.
     However, when comparing radiative and convective stars of the same
     \bv, they have almost the same hydrogen line strengths, but the
     convective star has a higher $T_{\rm eff}$.''  Simply put, H$\beta$ 
     and the \bv\ color index are both subject to changes brought on by
     atmospheric convection.  One therefore has to turn elsewhere, to 
     the UV spectrum according to BV82, to see this physical transformation.
     
     In our analysis of the broad-band {\it TD--1} fluxes, we follow
     basically the same approach taken in BV82, i.e., for each individual
     star we use the ratio of the broad-band 1965\AA\ flux to the visible
     flux in order to impute a $T_{\rm eff}$.
     The difference is that we stay entirely within the observational plane 
     instead of transforming the flux ratio (or equivalently the ultraviolet 
     minus visible color index) into an explicit value of $T_{\rm eff}$.  
     We write, \csxfv\ $\equiv m$(1965\AA) -- $V$, where  the UV magnitude 
     is given as usual by $m(\lambda) = -2.5\log f_{\lambda} - 21.10$, and 
     $f_{\lambda} = f_{1965}$ is the tabulated {\it TD--1} flux.  No 
     adjustments have been made to the zero point of the {\it TD--1} flux scale,
     since Bohlin \& Holm (1984) found excellent agreement between the 
     {\it TD--1} and the IUESIPS flux scales near 1900 \AA.  The
     \csxfv\ color index computed from the broad-band 1965\AA\ flux of 
     each star is plotted against \bv\/ color in Figure 4.  For reference, 
     we provide a temperature scale along the right-hand side of the
     graph, which shows the conversion from UV color to $T_{\rm eff}$, 
     based on the fluxes predicted by ATLAS9 model atmospheres. 
     In this color--color plot, the visibility of different sequences
     of convective or radiative stars is limited only by the formal
     errors in the ultraviolet photometry, which are 0.1 mag or smaller.
     The corresponding uncertainty in the implied $T_{\rm eff}$
     is $\pm80$ K or less, which should be sufficient to resolve a 
     substantial temperature gap.\footnote{
     The stars farthest from the model atmosphere curve in Figure 4
     are HR 1104, HR 3113, and HR 3756.  HR 3113 is a bright giant
     (Lemke 1989), whose location far below the main
     sequence relation is the result of its high luminosity, as we
     show later on.  HR 1104 and HR 3756 are relatively bright at UV
     wavelengths, but {\it IUE} photometry fails to confirm their 
     {\it TD--1} photometry, suggesting that these stars are either highly
     time variable or else their {\it TD--1} photometry is in error,
     perhaps due to source confusion in the $11\arcmin \times 18\arcmin$
     entrance slot.}

     The range in \bv\/ colors plotted in Figure 4 includes the ''A--star
     gap.''  As in the previous figure, we see no indication in
     this plot for two distinct photometric sequences of stars, and no
     evidence for a drop in the numbers of stars in the color range
     $0.22 \leq \bv \leq 0.28$.  A histogram of the cumulative star
     count, presented in Figure 5, shows a smooth and steady progression  
     as a function of \bv\/ color, with no leveling off to indicate
     a decline in the number density of stars in the above color range.
     This result is only suggestive because the survey sample is roughly 
     flux limited, rather than volume limited.   Nevertheless, any selection
     effects in a flux-limited survey should be monotonic with \bv, 
     and thus unlikely to affect the detectability of a gap in the stellar 
     number density.

     For a more quantitative statistical assessment, we have
     examined the distribution of \csxfv\ as a function of \bv\/ color
     on either side of, and through, the domain of the ``A--star gap.''
     If the stars in any \bv\/ interval  congregate into two separate
     temperature sequences, the distribution in their \csxfv\ colors
     should appear bimodal.  However, in no \bv\/ color interval do we
     find this to be the case, nor do we find any evidence that the
     variation in the mean or median UV color as a function of \bv\/
     changes slope to reflect the possible presence of a ``plateau''
     among the A or F stars.  The mean value of \csxfv, computed for 0.01
     mag intervals in \bv, is illustrated in Figure 6.  According to a
     Kolmogorov-Smirnov (K--S) test, the trend in the average value of
     \csxfv,  from \bv = 0.15 to \bv = 0.33, is indistinguishable from
     a straight line.  The standard error in a linear least-squares
     fit is $\pm0.07$ mag; this corresponds to a full $1\sigma$ range
     in $T_{\rm eff}$ of 120 K,  which is a factor of four smaller
     than the vertical distance separating the two branches of stars
     found in BV82.

     Figure 6 also summarizes in box-plot format the change in the
     distribution of UV colors along the main sequence, for two \bv\/
     color bins blueward, one bin within, and two bins redward of the
     ``A--star gap.''  No plateau is apparent in the median value of \csxfv,
     or at any of the four other percentile levels depicted there.
     Our study of an extensive sample of broad-band {\it TD--1} 
     photometry thus fails to confirm the existence of the temperature
     bifurcation and temperature plateau previously suggested by BV82
     from her analysis of the narrow-band photometry.

     In a horizontally stratified atmosphere, the UV flux may emerge
     from a large range of heights, depending on the wavelength-- and
     temperature--sensitivity of the opacity (Nordlund \& Dravins 1990).
     Consequently, the broad-band UV photometry, perhaps
     more so than the narrow-band data, may average out astrophysically
     important wavelength-dependent effects due to convection andor
     photospheric granulation. Given this possible limitation of the
     broad-band {\it TD--1} data, as well as the generally
     low S/N of the narrow-band {\it TD--1} data, we have therefore 
     undertaken a survey of high quality UV spectra of a sample of 
     A-- and early--F stars with the {\it IUE} spacecraft. The results 
     of that effort are discussed in the next section.  They are shown 
     to confirm the conclusions we have reached here from the {\it TD--1} 
     photometry.

     \section{{\it IUE} Spectroscopy of A--F Stars}

     \subsection{Observations}

     The {\it IUE} spectra analyzed here consist of new observations
     as well as data from the public archives.  All of the spectra
     were obtained with the short-wavelength prime (SWP) camera in its
     low dispersion mode, and were taken through the large science
     aperture.  The spectral resolution is $\sim8$\AA.  A full
     description of the {\it IUE} spacecraft and the performance of
     the telescope shortly after launch can be found in Boggess et al.
     (1978).  Prior to our investigation, a number of {\it IUE} guest
     observers had obtained spectra of normal A-- and F--type stars.
     Most of those images were heavily overexposed at the long
     wavelength end of the SWP camera (by factors of 5 or more) in
     order to enhance the detection of high temperature chromospheric
     and transition region lines at wavelengths shortward of 1700\AA.
     However, a search of the archives yielded 32 spectra of 29 stars
     that were suitably exposed in the 1700--2000\AA\ region of interest
     to us, and those spectra are included here in our survey of
     A--F stars.  Nine of the same stars were reobserved by us.

     The new observations consist of 64 spectra of 63 stars, the large
     majority of which stars are classified as spectroscopically 
     normal.  We excluded chemically peculiar (CP) Ap stars, but did
     include a few metallic line stars since B\"ohm-Vitense (1981b)
     found no difference between the {\it IUE} spectra of Am stars and
     those of normal A stars at wavelengths between 1400\AA\ and
     2500\AA.  Our analysis confirms this for the 1700--2000\AA\ 
     region (see also Praderie 1969).
   
     Most of the new spectra were acquired during US\#2 shifts and
     were occasionally subjected to high background radiation conditions.
     However, our exposure times were very short, generally a few minutes
     or less (the longest being 8 minutes for 57 Tau), and so the 
     quality of the spectrum was never adversely affected
     by the background radiation level. The integration times were
     generally chosen to achieve a peak exposure of 180--210 DN at
     wavelengths near 1900\AA, to ensure that the exposure level would
     be close to the upper end of the linear section of the intensity
     transfer function (ITF).  Except in five cases, the spectrum was
     either trailed, with a single pass of the star across the large
     aperture, or else multiply exposed at three standard offset
     positions within the large aperture, creating a ``pseudo-trailed'' 
     spectrum.  Two observations were taken as double exposures at 
     different positions within the large aperture.  Three others 
     were single exposures made at the normal location and were 
     subsequently processed as point source spectra.

     The stars included in this survey, for which we have either 
     new or archival spectra, are listed in Table 1.  The table
     provides ($V$, \bv) on the Johnson system, plus $uvby\beta$
     photometry gathered from the Hauck \& Mermilliod (1990) catalog
     or from information cited in the SIMBAD database; 
     estimates of the effective 
     temperature, $T_{\rm eff}$, derived from $uvby\beta$ using 
     the procedures described by Napiwotzki et al. (1993), which 
     incorporate improvements to Moon \& Dworetsky (1985); and comments 
     that flag the known binaries and $\delta$ Scuti variables in our 
     sample (Breger 1979, L\'opez de Coca et al. 1990).  As was true
     of the {\it TD--1} sample discussed earlier, the $T_{\rm eff}$
     values derived here show no evidence for a temperature plateau
     or for separate temperature sequences, and the rms spread
     of $\pm120$ K at a given \bv\/ or $b - y$ color index is not
     unexpected in light of either the calibration uncertainties cited
     by Napiwotzki et al. (1993) or the typical errors in $uvby\beta$ 
     photometry quoted by Crawford \& Perry (1966) and Crawford (1975).

     Table 1 also lists the $\delta c_1$ and $\delta m_1$ indices for
     each star.  As defined by Crawford (1975), these quantities
     assume $\beta$ as the independent (temperature) variable:
     $\delta c_1(\beta)$ = $c_1$(obs) -- $c_1$(std) and $\delta
     m_1(\beta)$ = $m_1$(Hyades) -- $m_1$(obs).  Although not strictly
     orthogonal, $c_1$ and $m_1$ serve as luminosity (surface gravity)
     and metallicity parameters, respectively (the $m_1$ index becomes 
     progressively more sensitive to surface gravity at \bv $<0.1$:
     Moon \& Dworetsky 1985).  The fundamental main sequence
     relations we adopt here are those established by Crawford \&
     Perry (1966) and Crawford (1975, 1979), as updated and extended
     by Hilditch et al. (1983).  In our analysis we explicitly allow
     for the fact that the $m_1$ relation is based on the Hyades Cluster,
     which is metal-rich compared to the Sun (Cayrel, Cayrel de
     Strobel, \& Campbell 1985).

     From each spectrum we measured an average flux at seven
     different wavelength positions from 1790\AA\ to 1960\AA.  The
     last bin, centered on 1960\AA, falls at the end of the SWP camera
     order. These mean fluxes were binned over wavelength intervals
     that were either 20\AA\ or 40\AA\ in width.  The positions of
     the wavelength bins are illustrated in Figure 7, where they are
     drawn as horizontal lines beneath a spectrum of $\alpha$ Aquilae
     (Altair).  The central wavelengths and widths of the bins
     were chosen to avoid known camera artifacts (the reseau marks)
     and also, to the extent possible, the strongest absorption
     features in the stellar spectrum.

     The average fluxes measured in three representative wavelength
     bins are compiled in Table 2.  We also give the corresponding
     Ultraviolet minus Visible colors, $[\lambda] = m(\lambda) - V$,
     where $m(\lambda)$ is the UV magnitude defined as before.  The
     spectral images listed in Table 2 have all been extracted and
     calibrated with the NEWSIPS pipeline processing system (Nichols
     et al. 1994).  As compared to the original IUESIPS
     processing scheme (Harris \& Sonneborn 1987), NEWSIPS provides a
     much more optimal and uniform extraction, as well as automatic
     corrections for a variety of wavelength--, temperature--, and
     time--dependent instrumental effects.  Before the NEWSIPS
     spectra were made available, however, we also measured the
     same images independently with the standard IUESIPS processing,
     except that we separately applied a number of corrections and
     adjustments. These were intended to compensate for:  
     (1) the difference between the actual trail length and the nominal 
     trail length assumed by IUESIPS (Garhart 1992a); (2) the quantization 
     of the clock cycle of {\it IUE\/}'s on-board computer, which affects 
     exposure segments shorter than $60^{\rm s}$ (Harris \& Sonneborn 
     1987, Oliversen 1991); (3) the temperature-sensitivity of the 
     camera response, which was generally a $<1$\% effect (in no case 
     was it $>2\%$); and (4) the systematic on-orbit decay in the 
     sensitivity of the SWP camera since the date {\it IUE} was launched 
     (Bohlin \& Grillmair 1988; Garhart 1992b).  In the final step we 
     used a table of sensitivity corrections kindly provided to us by 
     R. Bohlin (1995, priv. comm.), which served to update the information 
     published in Bohlin \& Grillmair (1988).  As described in the next 
     paragraph, these IUESIPS corrected fluxes are on the whole 5--7\% 
     brighter than the corresponding NEWSIPS measurements.  There is a 
     similar zero point offset in the latest HST/FOS flux scale with
     respect to the NEWSIPS calibration (Colina \& Bohlin 1994; Bohlin
     1996), which suggests that our IUESIPS corrected fluxes are very
     nearly on the FOS scale.  

     To enable a conversion from the NEWSIPS flux scale to the IUESIPS
     scale, Table 2 includes a mean correction factor for each star,
     $C_{\rm IUE}$, which is an average over the full set of seven 
     wavelength bands. Table 3 provides a mean correction factor for 
     each wavelength bin, averaged over the entire set of stars.  Both
     corrections are expressed in the sense of:  the magnitudes by
     which the IUESIPS spectral fluxes are {\it brighter} than the
     NEWSIPS values listed in Table 2.  As a general rule, the reader
     should be able to recover very nearly the actual IUESIPS measurements
     by combining the star-by-star and the bin-by-bin corrections. 
     Because the NEWSIPS fluxes provide a marginally better fit
     to theoretical model atmosphere fluxes, as we will describe further
     below, there may be slight preference for the NEWSIPS fluxes over
     the IUESIPS ones.  However, the choice of processing scheme has 
     little or no impact on the conclusions that we reach here (this
     will be made more evident, e.g., in Figure 14 below).

     \subsection{Color--Color Plots}

     A plot of [$\lambda$], the UV color index from Table 2,
     versus the \bv\/ color
     index is shown for wavelengths of 1810\AA\ and 1960\AA\ in Figures
     8a and 8b, respectively.  Drawn for comparison in each panel is the
     main sequence relation predicted by ATLAS9 models (Kurucz 1993a,b,
     1994), as well as a reddening line corresponding to a
     color excess $E(\bv) = 0.1$.  The ATLAS models assume a surface
     gravity of $\log g = 4$, solar metal abundances, a microturbulent
     velocity $\xi_{\rm turb} = 2$ km s$^{-1}$, and, in the models that
     initiate convection (at $T_{\rm eff} < 9000$ K), a mixing-length to
     scale-height ratio equal
     to $l/H_{\rm p} = 1.25$.  Also displayed are a pair of models at
     lower surface gravity, $\log g = 3.5$ and $\log g = 3$, and a pair
     with higher metal abundances, [M/H] = 0.3 and [M/H] = 0.5 (where 
     the usual bracket notation specifies the logarithmic difference
     in M/H with respect to the Sun).  The surface gravity that defines
     the main sequence relation is a compromise between a higher value
     that gives the best fit to the {\it IUE} photometry at 1960\AA\ 
     and a lower value required at 1810\AA.  In both panels, and for
     the other five wavelengths not shown here (the appearance of all
     these other plots is similar to Figure 8), the observations cluster around 
     the model atmosphere curve.  Neither panel of Figure 8 shows any 
     sign of a temperature plateau or of two distinct temperature sequences.
     The vertical scatter in UV color around the model atmosphere
     relation is substantial, amounting to a range  of $\sim0.65$
     mags at a given \bv.  A scatter this large far exceeds the
     photometric errors of {\it IUE}, which are 3.5\% or less in broad
     photometric bands (e.g., Garhart \& Nichols 1995).  We have
     verified this error estimate for our own particular wavelength 
     bands by reprocessing more than 50 SWP low dispersion spectra
     of the {\it IUE} calibration standard BD+$28^{\circ}$4211, which
     cover the same time period as our A--F star spectra.  The
     resulting flux measurements are very nearly Gaussian distributed
     with a $\sigma$ = 0.035 mags.

     Given the slope of the normal reddening line, the better part of
     the scatter we observe among the A--F stars in Figure 8 cannot be
     due to interstellar extinction, since it is clear that
     differential reddening moves a star mostly parallel to the main
     sequence rather than across it.  Furthermore, all of our stars are
     located close enough to the Sun that reddening effects are generally
     negligible, E(\bv) $<0.02$ (Perry \& Johnston 1982). It is therefore 
     difficult to see how a plausible amount of extinction could
     displace a star from a reasonable point of origin on the main
     sequence, somewhere to the left, to a location either well above
     or well below this curve.  Some stars, e.g., $\beta$ Pic, may suffer 
     appreciable local or circumstellar extinction (e.g., $A_v = 0.19$ 
     for $\beta$ Pic, according to Lanz, Heap, \& Hubeny 1995), but their
     circumstellar particles are thought to be large enough that they 
     should produce very little selective extinction, and hence should 
     not displace a star off the main sequence relation in a color--color 
     plot like Figure 8.  Switching the abscissa from \bv\ to an alternative
     photometric index, e.g., $\beta$ or $b - y$, leads to some 
     rearrangement of the points for individual stars (see Figure 9),  
     but neither eliminates nor reduces the amount of scatter.

     \subsection{Instrumental Sources of the Scatter}

     Among possible instrumental origins of the scatter in Figures
     8 and 9, we mention the following:  (1) the observing and image
     processing mode (trailed, multiple-exposure, or point source),
     (2) the $\beta$ angle of the observation, i.e., the orientation
     of the star with respect to the direction of the Sun at the time
     of observation, (3) the focus setting of the telescope, (4) the
     DN (or data number) exposure level or the exposure time of the 
     spectral image, and (5) anomalies in the pointing and tracking 
     of the telescope during an observation.

     We consider first the observing and processing mode.  Figure 10a
     plots the vertical deviation of the \csxoh\ color of each star from
     the model atmosphere curve, $\delta$1960 $\equiv$ \csxoh$_{obs}$ --
     \csxoh$_{model}$. The scatter does not appear to depend in any
     systematic fashion on the imaging mode, whether the image was 
     trailed, pseudo-trailed (multiply exposed), or acquired in point 
     source mode. At any given \bv, points for the different modes are 
     intermixed. Likewise, in Figure 10b there is no correlation between 
     $\delta$1960 and the $\beta$ angle of the observation over the full 
     range of angles observed, from $40^{\circ}$ up to $105^{\circ}$.  
     This effectively rules out any significant contamination of our 
     photometry by scattered solar radiation within the telescope. 
     For the same reason, the spurious streak of light in the Fine 
     Error Sensor (FES) formed by this light leak must never have 
     compromised the guiding accuracy of our trailed spectra.

     There is also no apparent relationship between the color
     deviations and the focus setting of the telescope (Figure 10c),
     the exposure level (DN) of the observation (Figure 10d), 
     or the exposure time of the observation (Figure 10e).
     Residual non-linearities in the ITF therefore can be eliminated 
     as a major cause of the scatter. Finally, we have confirmed by
     inspection of the observing log sheets that there were no 
     significant problems with the tracking or pointing of the telescope
     in the course of any observation, except that of 49 UMa (image
     SWP46683), which the telescope operator noted as having a trail
     error of $\sim4\arcsec$. The fluxes measured from this spectrum
     are not appreciably offset from the model atmosphere curve.

     \subsection{Intrinsic Causes of the Scatter:  The Role of Stellar
     Rotation}

     If the onset of convection among the A--F stars causes the scatter
     observed in two--color plots such as Figure 8, then it is worthwhile
     inquiring whether this scatter correlates with other established
     proxies for convection, such as chromospheric emission
     in the \ion{C}{2} 1335\AA\ and \ion{H}{1} Lyman $\alpha$ lines.
     We have demonstrated elsewhere that the strengths of these 
     chromospheric features vary widely in A--F star spectra (Simon \&
     Landsman 1991; Landsman \& Simon 1993; Simon, Landsman, \& Gilliland
     1994), which suggests there is a broad range in the convection 
     zone properties of these stars.  However, the deviations in UV  
     colors plotted in Figure 8 show no correlation with the normalized
     \ion{C}{2} fluxes listed by Simon \& Landsman (1991), and plots of
     $\delta$1810 as well as $\delta$1960 against X-ray emission (from 
     Simon, Drake, \& Kim 1995) are merely scatter diagrams.  Thus, we 
     find no direct link to connect the UV colors of the A--F stars with 
     chromospheric and coronal proxies for convection.

     The question of whether the stellar rotation speed regulates the
     onset of convection, as originally conceived by B\"ohm-Vitense \& 
     Canterna (1974) and BV82, is also susceptible to direct test.  
     This involves a comparison of $\delta$1810 and $\delta$1960 with 
     published values of $v \sin i$.  Such a comparison indicates 
     that the deviations in UV colors are uncorrelated with the stellar 
     rotation rate, for stars inside the ``A--star gap,'' $0.22 \leq
     \bv \leq 0.28$, and also for those outside it.  Consider Figure 11.
     If stars within the gap are typically faster rotators, and if they 
     define a low temperature radiative branch, characterized by generally
     larger UV color indices, then their points in this figure should fall
     systematically below those of the more slowly rotating stars that
     form the higher temperature convective plateau. Several of the gap 
     stars, notably the $\delta$ Scuti variable 71 Tau, may follow
     this pattern, but there does not appear to be any distinct segregation
     of the points in either panel of Figure 11.

     In addition, a number of stars inside the gap appear to be 
     moderately slow rotators; some of them exhibit either chromospheric 
     emission in \ion{C}{2} or else have been detected in X rays (e.g., 
     22 Boo and $\epsilon$ Cep).  If these stars are intrinsically slow 
     rotators that have turned convective, the question is, why are they 
     inside the gap instead of outside it, at $\bv > 0.28$?  And if they 
     are intrinsically rapid rotators seen nearly pole-on, why have they
     not remained radiative?  Either way, they stand in conflict with 
     the idea that rotation regulates the onset of convection among the 
     A--F stars.  Moreover, as noted previously by Simon, Drake, \& Kim
     (1995), there are some rapidly rotating stars inside the gap showing
     both UV chromospheric emission and coronal X-ray emission, e.g., Altair
     and $\alpha$ Cep, which flatly contradict the idea that rapid rotation
     acts to suppress the onset of convection within the ``A--star gap.''

     \subsection{The Effect of Luminosity and Metallicity}

     In a sample of A--F stars chosen as broadly as ours, it would 
     not be unduly surprising to find a range in the luminosity 
     (=surface gravity) and chemical composition of the individual 
     stars (cf. Berthet 1990, Edvardsson et al. 1993).  Some impression
     of the importance of these parameters to the overall scatter
     in brightness at UV wavelengths can be derived from the model
     atmosphere fluxes plotted in Figure 8.  With only a moderate
     range in the surface gravity, from $\log g = 4$ to $\log g = 3$,
     and in the metals to hydrogen ratio from solar to three times
     solar, that is, [M/H] = 0 to [M/H] = 0.5, the locus of model colors
     moves transverse to the standard main sequence and manages to cover
     virtually the entire range of UV and optical colors we have
     observed.  This suggests that the observed scatter in the 
     color--color plots might be accounted for by a modest spread
     in these properties throughout our sample of A and F stars.  
     In fact, a plot of $\delta$1960 shows strong trends with both
     the Str\"omgren luminosity index $\delta c_1$, and with
     the metallicity and line-blanketing index $\delta m_1$.

     We have devised a method to adjust the observed colors of each star
     for potential gravity and metallicity effects, which makes use
     of the $c_1$ and $m_1$ indices. In effect, we attempt to restore 
     each star back to the location it would have on the standard main 
     sequence for solar abundances and a surface gravity of $\log g = 4$.  
     Specifically, we define the correction for the UV color index at
     1960\AA\ (and similarly for all the other UV wavelengths) by a pair
     of linear relations,
          $$\delta 1960 = \phi_1(T_{\rm eff}\,|\,\log g)\delta c_1,$$
          $$\delta 1960 = \phi_2(T_{\rm eff}\,|\,M/H)\delta m_1,$$
     where $\phi_{1,2}$ are polynomials in the $\beta$ and/or $b - y$ index
     of integral and quarter-integral order.  The coefficients in these
     polynomials were determined from a least-squares fit to the output 
     of the ATLAS9 model atmosphere grid over the range $3 \leq \log g 
     \leq 4$ and $-0.5 \leq [M/H] \leq 0.5$.  The Str\"omgren indices, 
     $\delta c_1$ and $\delta m_1$, are the observed values listed in Table 
     1. Similar expressions can be devised for the \bv\/ color index, but
     since the photometric indices of the Str\"omgren system are much
     more precisely defined and calibrated in terms of stellar properties,
     we prefer to work with $b - y$ instead.  Thus,
          $$\delta (b-y) = \psi_1(T_{\rm eff}\,|\,\log g)\delta c_1,$$
          $$\delta (b-y) = \psi_2(T_{\rm eff}\,|\,M/H)\delta m_1.$$
     Here, $\psi_{1,2}$ are also least-squares fitted polynomials in
     the $\beta$ index and/or $b - y$.  As noted earlier, the fiducial 
     relation for $m_1$ is based on the Hyades Cluster, and so we 
     adjust for the fact that $\delta m_1 = 0$ refers to a metal-rich 
     composition (cf. Berthet 1990, Smalley \& Dworetsky 1995).

     A modified color--color plot using $b - y$ and UV colors corrected
     as just described is presented in Figure 12.  To indicate that
     the colors have now been adjusted, we attach double primes
     to both indices, $(b - y)''$ and \csxoh$''$.  The result of
     having made these corrections is to produce a much tighter fit to
     the model atmosphere main sequence relation and to leave a smaller
     scatter in the observed color--color relation.  In the process,
     these adjustments also eliminate the previously found correlated
     dependences of $\delta$1960 on both $\delta c_1$ and $\delta m_1$.  

     With respect to the model atmosphere sequence,  the rms scatter
     at 1960\AA\ drops in half from $\pm0.25$ mags to $\pm0.13$ mags.
     Despite the reduced amount of scatter, the points still 
     do not form a plateau or line up into two distinct sequences of 
     stars, and furthermore a plot of the UV color deviations against 
     rotation speed shows, as before, no obvious trend with $v \sin i$
     (Figure 13).  We note that some of the hottest stars in our sample,
     at $T_{\rm eff} >  8000$ K, are slightly displaced to more negative 
     \csxoh$''$ colors with respect to the radiative ATLAS9 models,  
     but the difference is at most 0.2 mags.  Except for this slight
     offset in the NEWSIPS colors, we can point to no other systematic
     trends in the scatter in our color--color plots following the
     luminosity and abundance adjustments we have made. If, for example,
     we reconsider the question of instrumental scatter raised in Section
     3.3, no correlation emerges in the adjusted data with respect to
     (1) the imaging mode, (2) the $\beta$ angle of the observation, 
     (3) the camera focus setting, or (4) the DN exposure level of 
     the spectrum. Figure 10 offers the details.

     \subsection{Microturbulence}
     
     Our use of ATLAS models to calibrate $\delta m_1$ against [M/H] 
     makes the implicit assumption that the Str\"omgren index is a
     surrogate for the chemical composition of a star.  The fact that 
     the adjusted colors we derive show less scatter, rather than 
     more, tends to validate this hypothesis.  However, the effects
     of a range in metallicity on model atmosphere fluxes can be
     mimicked by a spread in microturbulent velocity (for a related
     discussion, refer to Smalley 1993).  An increase in the overall
     metal abundance by a factor of 2 above solar raises the predicted 
     $m_1$ index by nearly 0.02 mags.  An increase in $\xi_{\rm turb}$ from 
     2 km s$^{-1}$, the nominal value we have adopted here, to 4 km 
     s$^{-1}$ does precisely the same.  At wavelengths in the {\it IUE}
     range, metallicity has the stronger influence of the two.  For a 
     factor of two higher metallicity, \csxoh\ changes by 0.4 mags;
     but doubling the value of $\xi_{\rm turb}$ from 2 km s$^{-1}$ to 
     4 km s$^{-1}$ produces only a 0.1 mag change. Consequently, if the
     $\sim0.65$ mag scatter observed in the colors of the A stars is due
     to the influence of microturbulence alone, then the range in 
     $\xi_{\rm turb}$ in our {\it IUE} sample would have to go
     well outside the limits of 2--4 km s$^{-1}$ that have
     been established from studies of line profiles and abundance 
     analyses for stars of this class (e.g., Lane \& Lester 1987; 
     Lemke 1989; Coupry \& Burkhart 1992). 

     Some of the residual scatter of $\pm0.13$ mags that remains in 
     color--color plots like Figure 12 could be associated with a 
     normal spread in the individual values of $\xi_{\rm turb}$, but 
     to prove this would require velocity estimates for many more stars 
     in our {\it IUE} sample than are now available. There is some 
     overlap between our sample and the F-star sample of Balachandran 
     (1990).  Her study provides microturbulent velocities (albeit only
     indirect estimates) for a half-dozen stars in common, including
     HR 2740, which has the most
     extreme color deviation $\delta 1960''$ of any star that we observed.  
     The microturbulent velocity assigned by Balachandran to HR 2740 is 
     unexceptional, $\xi_{\rm turb} = 1.9$ km s$^{-1}$.  The values  
     she reports for the remaining stars are also in the range of 1.8 
     to 2.1 km s$^{-1}$, being equally unremarkable.  Five other
     stars in our survey were analyzed by Russell (1995), who obtained
     microturbulent velocities $\xi_{\rm turb} \approx 2.5$ km s$^{-1}$
     for each of them; the range in $\delta 1960''$ is 0.18 mags.
     On the basis of these two studies, then, we would be inclined to
     say that microturbulence likely has only a minimal impact on the
     spread in UV colors of the A--F stars, but at this stage our
     assessment is by no means definitive.

     \subsection{Intrinsic Variability}

     Unresolved binaries have been suggested as a major source of
     ``noise'' in $uvby\beta$ photometry (Crawford \& Perry 1966). 
     When it comes comes to explaining the scatter in the UV colors 
     of A stars, however, binary status seems to matter very little.
     There is no difference in the range of $\delta 1960''$ among
     the single stars and those classified as either spectroscopic 
     or wide doubles (see Figure 13b).  Also, the rms scatter for 
     the binaries, $\pm0.12$ mags after the luminosity and metallicity
     adjustments described in Section 3.5, is very close to the rms
     scatter of $\pm0.13$ mags in the sample as a whole.

     Similarly, the behavior of the known $\delta$ Sct pulsators in
     our sample is indistinguishable from that of the non-$\delta$ Sct
     stars.  The rms deviation in the adjusted colors of 
     the variables, $\pm0.12$ mags, is practically the same as that
     of the full sample.  Moreover, a histogram of the individual
     deviations, $\delta 1960''$, shows a similar, if not slightly 
     narrower, spread for the variable stars than for the non-variables 
     (Figure 14). The largest amplitude in $V$ among all the $\delta$ 
     Sct stars we observed is 0.05 mags for $\upsilon$ UMa (L\'opez de 
     Coca et al. 1990). For the remainder, the amplitudes at $V$ are 
     typically $\leq0.02$ mags. Three of the $\delta$ Sct stars we observed 
     had spectra in the archives:  LW Vel (HR 4017) was virtually 
     the same in 1800--2000\AA\ brightness in its archival spectrum 
     as when we observed it (to within 4\%); $\rho$ Tau was 10\% 
     brighter in our spectrum; and 71 Tau was 20\% brighter in our 
     observation than it was five months later when Monier \& Kreidl
     (1994) observed it.  During the single 8-hour {\it IUE} shift 
     that 71 Tau was observed by those authors, no variations were seen 
     in the UV brightness of this star on a timescale comparable with 
     its $\sim$3.9-hour stellar pulsation period.\footnote{
     71 Tau lies below the standard sequence in our color--color
     plots.  It has a late-type companion and, given its \bv\ color,
     might have been expected to appear comparatively brighter in 
     the UV, not fainter.}

     In the course of our survey, we also observed HD 224638, a 
     member of the $\gamma$ Doradus class of non-radial $g$-mode 
     pulsators (Balona et al. 1994; Mantegazza et al. 1994). 
     The fluxes measured from the {\it IUE} spectrum of HD 224638 
     show only a small departure of --0.05 mags from the standard 
     model atmosphere color--color sequence.

     The majority of the non-variable stars observed with {\it IUE} 
     lie within the boundaries of the well-known pulsation instability 
     strip (e.g., Breger 1979).  Presumably, under the right
     circumstances, these stars might also be subject to pulsational 
     instability. However, if their pulsation amplitudes were small 
     enough, photometric variability might be noticeable only at UV 
     wavelengths, where the stellar spectral energy distribution is 
     several times more sensitive to $T_{\rm eff}$ than it is at 
     visible wavelengths. If we assume for discussion that the residual 
     rms scatter of $\pm0.13$ mags in the UV colors is the result of 
     intrinsic variability, the implied temperature variation amounts 
     to $\Delta T_{\rm eff} = \pm100$ K.  The corresponding brightness
     change expected at optical wavelengths for this $\Delta T_{\rm eff}$
     is $\Delta V = \pm0.07$ mags.  Variability this pronounced would 
     certainly have been noticed in time-series $uvby\beta$ photometry, 
     which routinely achieves an accuracy of just a few millimags.  
     On the other hand, variability at half that amplitude, 
     $\Delta V \approx \pm0.03$ mags, might well have been missed 
     in isolated photometry taken with no forethought of variability,
     especially if the changes happened to be irregular or multiperiodic 
     in character.

     Twelve of our stars had repeat measurements with {\it IUE} that 
     can be examined for ultraviolet variability.  Just two
     stars, 71 Tau and $\rho$ Tau, both $\delta$ Sct variables, showed 
     clear signs of variability in excess of the mean photometric
     errors of $\pm0.04$ mags. In the case of 71 Tau, the timescale 
     of the UV variability is unknown but the changes appear to be
     unrelated to the short-period $\delta$ Sct 
     pulsation because of (a) the lack of variability reported in the
     prior study of Monier \& Kreidl (1994), and (b) the very
     improbable 10:1 UV-to-optical amplitude ratio implied by the
     color deviations of the {\it IUE} archival spectra of this star.
     Thus, while we cannot show from the observational evidence, such 
     as it is, that intrinsic variability makes a significant contribution 
     to the scatter in our two color plots, it may be premature to
     dismiss this possibility until the variations of a 
     star like 71 Tau are better understood and the full range of 
     its pulsational behavior has been more thoroughly explored.

     \section{Conclusions}

     We have used broad-band {\it TD--1} photometry and high quality spectra
     from the {\it IUE} spacecraft to explore the possible effects of
     subsurface convection on the spectral energy distributions of normal,
     main sequence A and F stars.  In particular, we have examined a proposal
     originally offered by B\"ohm-Vitense, that the onset of 
     convection among the A--F stars may lead to the formation of two
     distinct temperature sequences, one comprised of the convective
     stars, the other consisting of the radiative stars.  We see no evidence
     for any temperature bifurcation in the two samples of stars we have
     studied with {\it TD--1} and {\it IUE}.  Given the very small
     internal photometric errors of the {\it IUE} spectrophotometry,
     typically $\pm3.5$\% or better in observed flux, or equivalently
     $\pm25$ K in stellar effective temperature, we should have easily
     seen and thus confirmed a systematic separation in $T_{\rm eff}$ as
     large as the 500 K difference claimed by B\"ohm-Vitense from her
     study of the narrow-band {\it TD--1} photometry (BV82).    

     On the other hand, we do find a considerable dispersion in the UV colors
     observed with {\it IUE}, up to 0.6 mags or more at a given \bv\
     color index.  A number of possible extrinsic and intrinsic explanations
     for the scatter can be eliminated:  various instrumental and/or image
     processing effects, interstellar or circumstellar extinction, binarity,
     a spread in the rotation rates of individual stars, or the onset of
     convective activity in the form of hot chromospheres and coronas.
     The most plausible cause, we find, is a range in the surface gravities
     and metallicities of the stars in our survey. Through a comparison with 
     ATLAS9 model atmosphere fluxes, we can account for at least 0.4 mags of
     the scatter with a 1 dex spread in gravity and a 0.5 dex spread in the
     metals-to-hydrogen ratio.  Another 0.1--0.2 mags may arise out of a
     normal dispersion of a few km s$^{-1}$ in atmospheric microturbulence
     (through its influence on the degree of line blanketing).  A similar
     contribution may come from photometric variability if the stars in
     our survey prove to be marginally stable against pulsation.  Evidence
     for this is very weak.

     A more plausible interpretation for the residual UV scatter of 
     $\sim0.25$ mags can be found in the empirical relation between metal
     abundance and the Str\"omgren $\delta m_1$ index, which forms the 
     basis for the technique we have used to adjust the {\it IUE} colors for
     differing amounts of spectral line blanketing.  According to Figure
     6 of Berthet (1990), among the A and F stars there is a spread of
     0.4 dex or more in the Fe/H ratio at a fixed value of $\delta m_1$.
     Very nearly the same spread is present in the more homogeneous
     dataset of Smalley (1993: see his Figure 3).  A dispersion in [M/H] of
     this order overlaps the entire range of scatter shown here in Figure 8,
     and moreover is fully consistent with the size of the discrepancies
     between the spectroscopic and photometric determinations of [Fe/H]
     obtained by Balachandran (1990) for her set of F stars.  In our 
     analysis we also make a common simplifying assumption, which may not
     be truly justified, viz., that the chemical abundances of the elements
     responsible for most of the UV line blanketing scale together.
     In the 1700--2000 \AA\ region, \ion{Fe}{2} and \ion{Si}{1} both have
     a number of strong absorption lines in A--F star spectra.  In the
     spectroscopic survey of nearby F--G field stars conducted by Edvardsson
     et al. (1993), these two elements vary by 0.2 dex or more in relative
     abundance from one star to the next.  If this is true as well among the
     A--F stars, then a single photometric index like $\delta m_1$ can
     provide only a first-order correction for UV line blanketing; more 
     detailed abundance analyses may therefore be required in order to lower
     the scatter in our color--color diagrams below the current
     residual level of $\sim0.25$ mags.

     Although we find in our study no evidence for any gross changes in 
     the UV colors of A and F stars that might be associated with the
     onset of subsurface convection, the search for more subtle signatures
     in the photospheric lines and spectral energy distributions of these
     stars nevertheless remains a worthwhile goal.  In particular, the 
     correlation of photospheric features with established chromospheric
     and coronal proxies for convection would provide valuable constraints
     on both the proper model of convection, as well as the source of
     the non-radiative heating.  However, as we have shown in this paper,
     the detection of a photospheric signature of convection will first
     require a more detailed analysis of the effects of metallicity,
     gravity, and microtubulence --- and possibly variability --- on the
     spectra of the A and F stars.        
     
     We thank Dr. Y. Kondo and the staff of the {\it IUE} observatory for
     their assistance in the acquisition and reduction of these data.
     This research has made use of the {\small SIMBAD} database, operated
     by the Centre de Donn\'ees astronomiques de Strasbourg.  The authors
     thank R. L. Kurucz for discussions and assistance with the ATLAS
     model atmospheres, and R. C. Bohlin for providing us with updates
     to the {\it IUE} camera sensitivity function.

     \newpage
      
     \begin{figure}
      \epsscale{0.85}
     \plotone{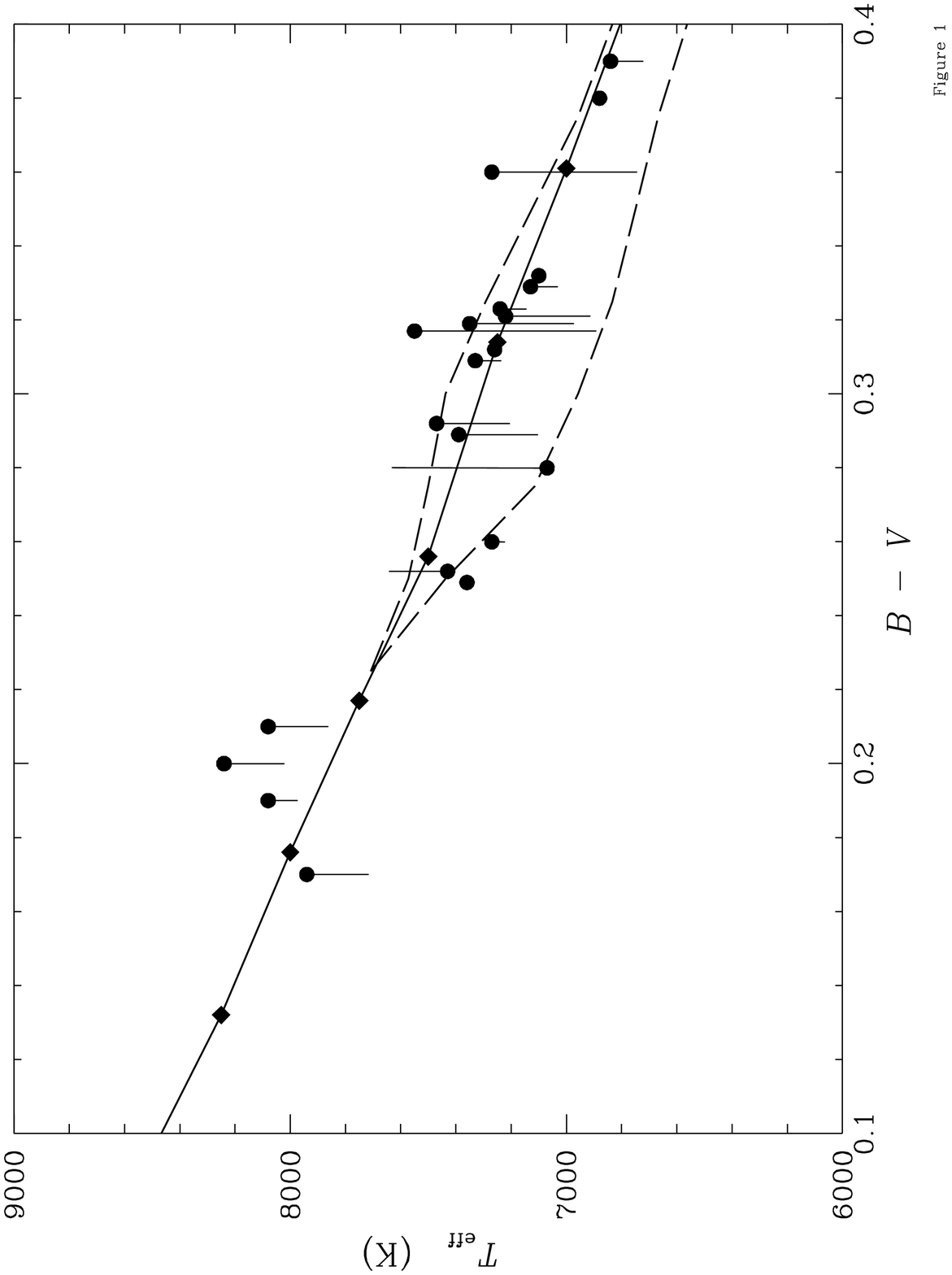}
     \figcaption {Effective temperature versus \bv\ color for main sequence
	A--F stars.  The filled circles are estimates derived from {\it TD--1}
	photometry by B\"ohm-Vitense (1982, Table 3), while the vertical lines
    	are the revisions we obtain by applying the method of Napiwotzki
        et al. (1993) to $uvby\beta$ photometry.  The diamonds connected
	by solid lines are ATLAS model atmospheres (Kurucz 1993a).  The dashed
        lines denote the upper convective branch and lower radiative branch
        of B\"ohm-Vitense (1982), as transcribed from Figure 7 of that paper, 
	except that we omit the evolved giants and bright giants that were 
	included in the original figure.}
     \end{figure}
     \begin{figure}
      \epsscale{0.95}
     \plotone{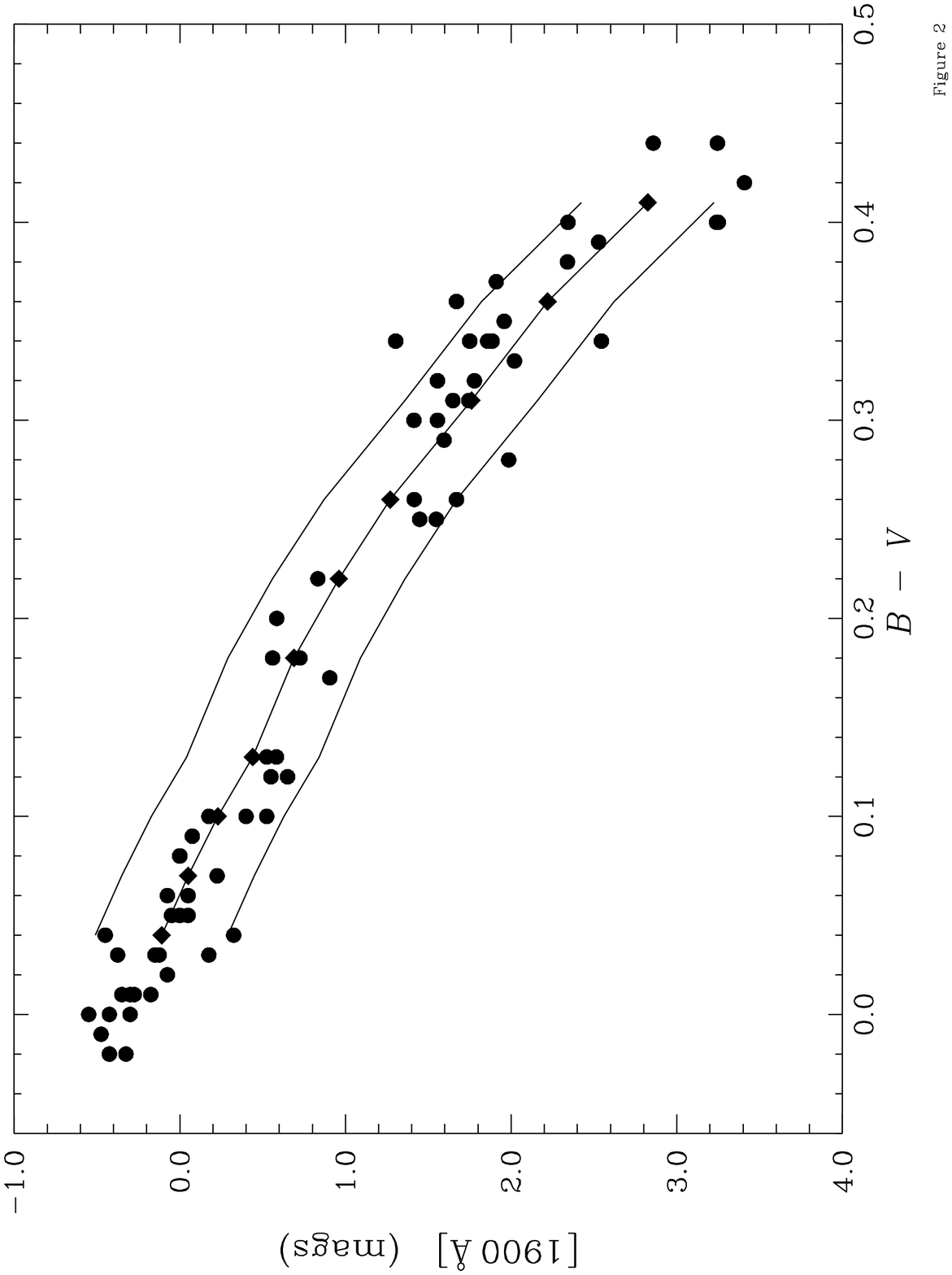}
     \figcaption {Narrow-band {\it TD--1} photometry for A--F stars ({\it
	filled circles}) and model atmosphere fluxes ({\it filled diamonds}).
        $m$1900\AA) is the narrow-band {\it TD--1} magnitude.  Plotted here
        are the stars that appear in Tables 1 and 2 of B\"ohm-Vitense (1982).
        The solid lines drawn parallel to the model atmosphere curve are
        offset vertically by $\pm0.4$ mags.  }
     \end{figure}
     \begin{figure}
     \plotone{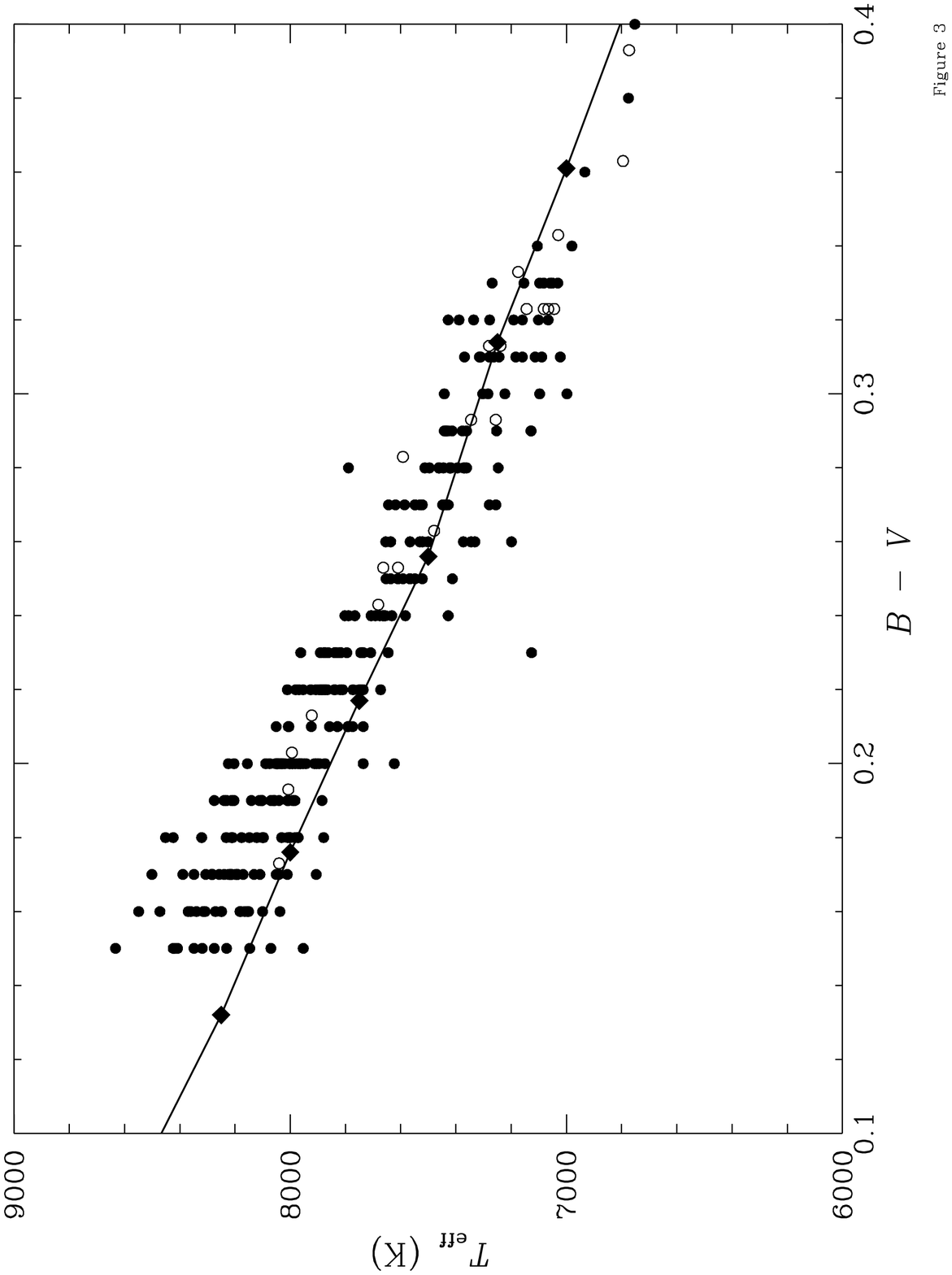}
     \figcaption {Effective temperatures from Str\"omgren photometry for
	A--F stars with {\it TD--1} broad-band photometry.  Chemically
	normal stars are shown by filled circles, stars from 
	B\"ohm-Vitnese (1982) as open circles (offset slightly in
	\bv\ color for clarity).}
     \end{figure}
     \begin{figure}
     \plotone{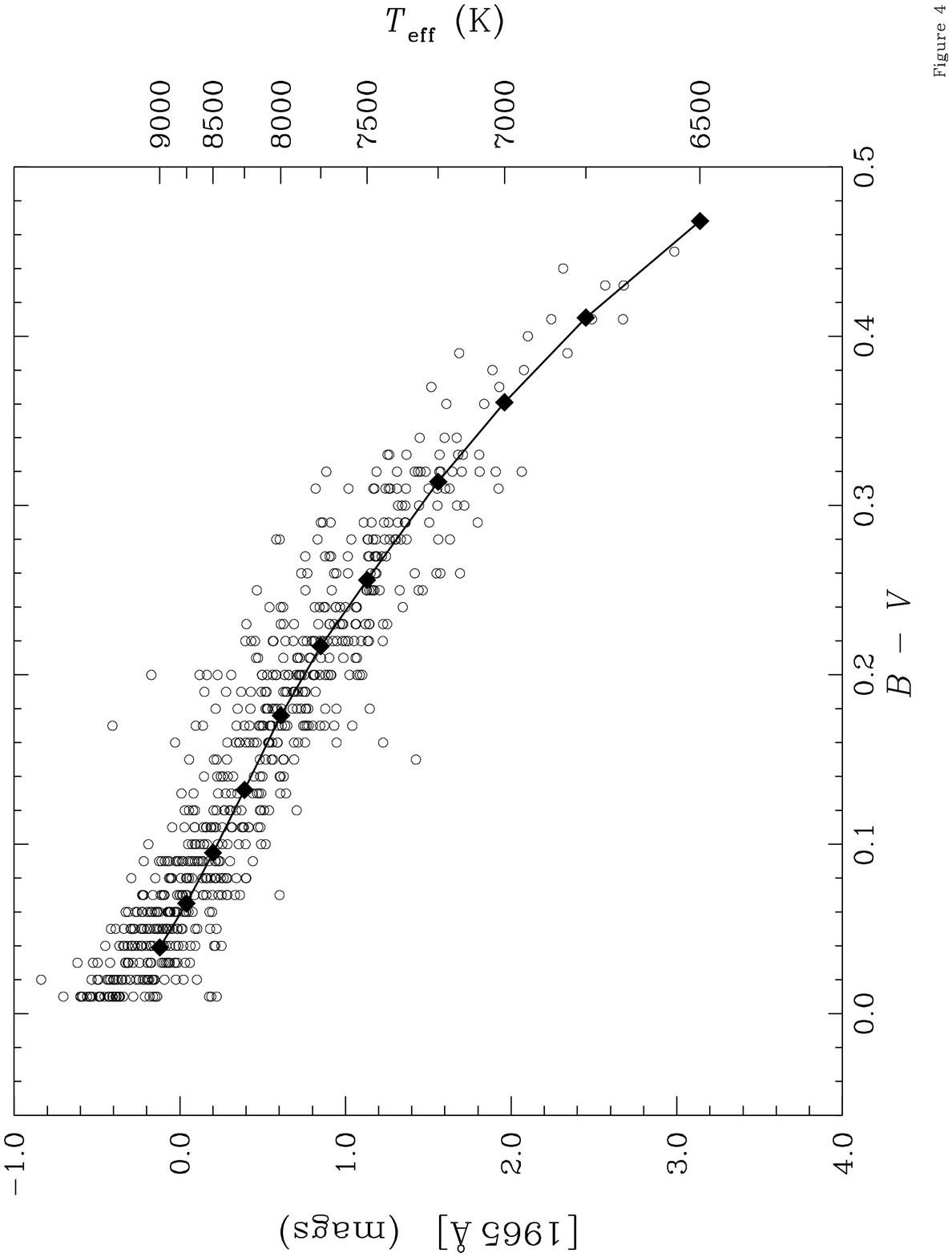}
     \figcaption {Broad-band ultraviolet (1965\AA) minus $V$ color for
	{\it TD--1} stars, plotted versus \bv\ color.  The diamond symbols
	denote ATLAS9 models.}
     \end{figure}
     \begin{figure}
     \epsscale{0.95}
     \plotone{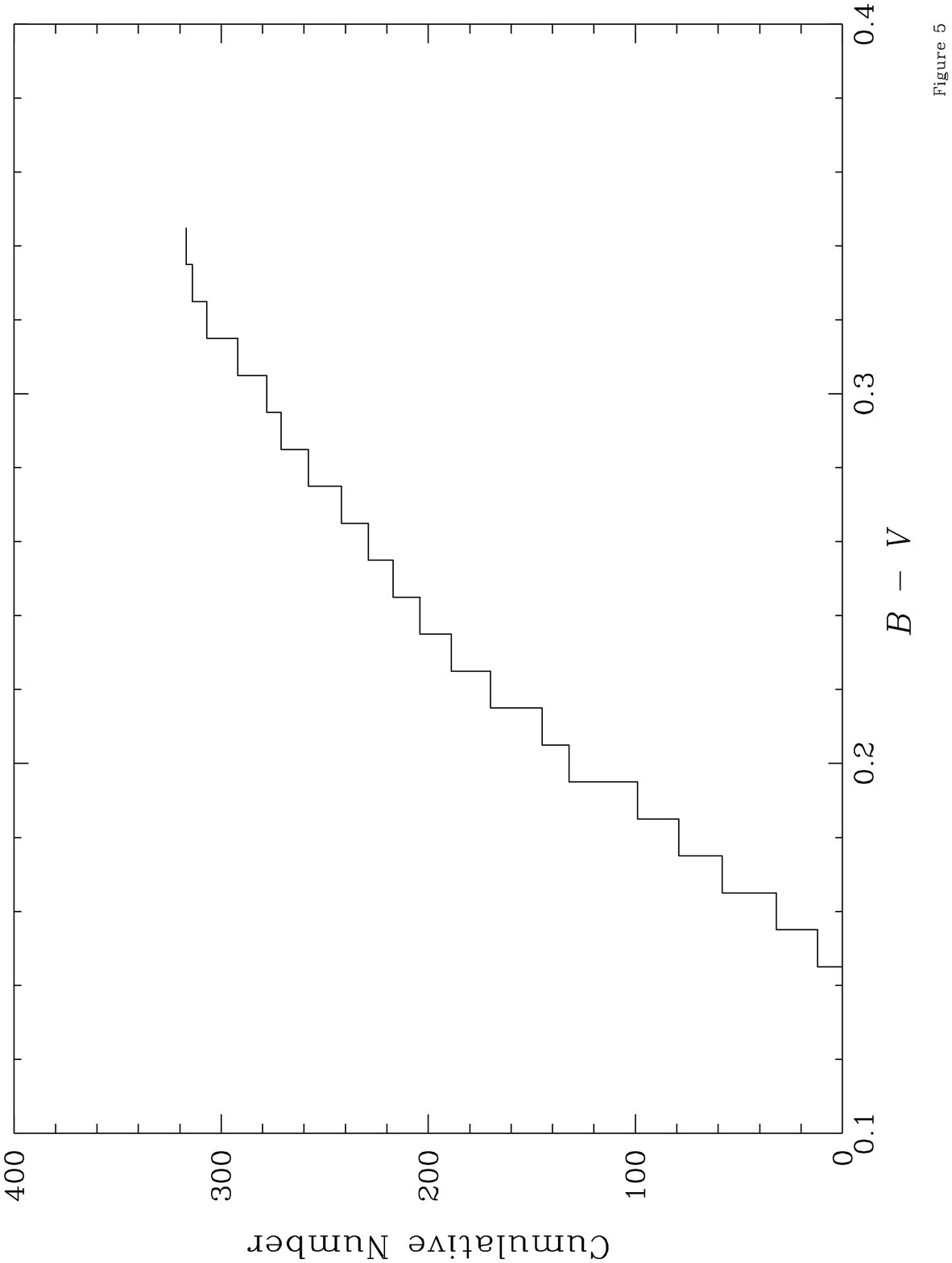}
     \figcaption {Cumulative count of {\it TD--1} stars with colors in
	the range $0.15 \leq \bv \leq 0.33$. An  
	absence of stars in the vicinity of the A--star gap, $0.22 \leq
	\bv \leq 0.28$ would be signified by a horizontal plateau in
	the cumulative distribution function.  We see no evidence in this
	range of colors for a gap in our {\it TD--1} sample.}
     \end{figure}
     \begin{figure}
     \epsscale{0.95}
     \plotone{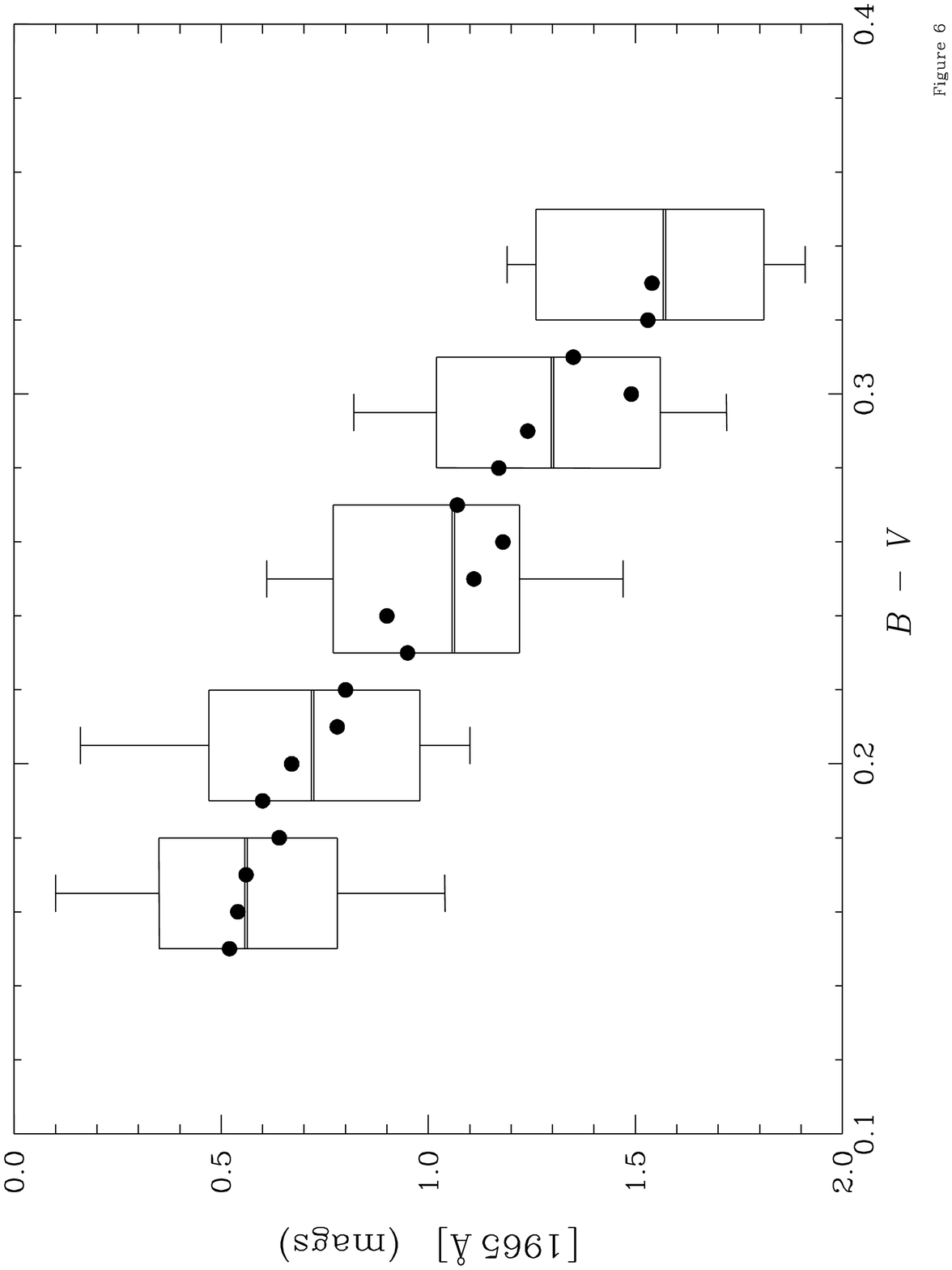}	
     \figcaption {The distribution function of broad-band 1965\AA\
	{\it TD--1} colors of main sequence stars for five \bv\ color
	bins.  The percentile levels shown by the box plots are at:  
	95\% and 5\% ({\it vertical bars}), 84\% and 16\% ({\it upper
	and lower edges of the box}), and 50\% ({\it double bar within
	the box}).  Also shown is the mean \bv\ color, computed for
	0.01 mag intervals in \bv.}
     \end{figure}
     \begin{figure}
     \epsscale{0.95}
     \plotone{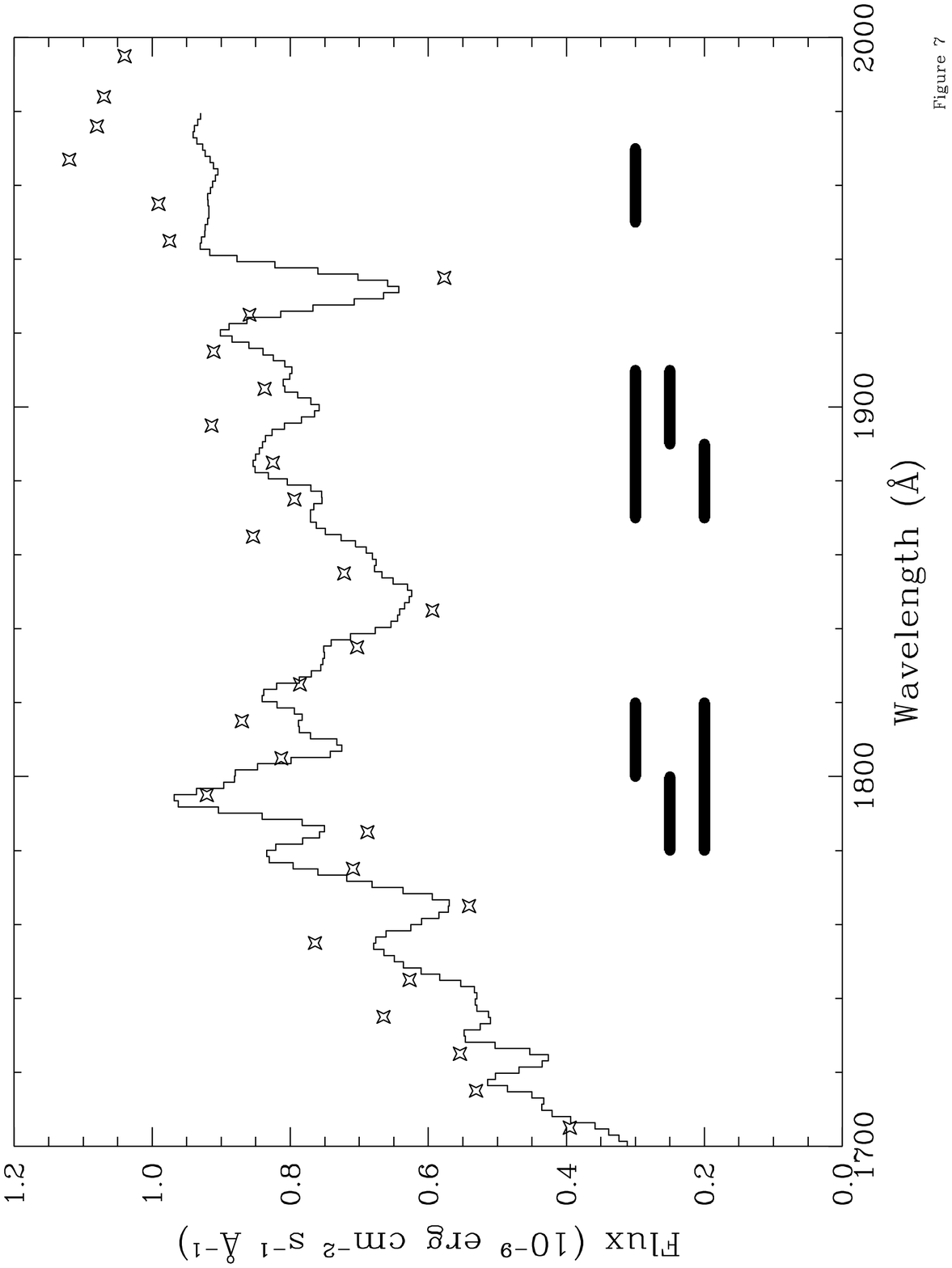}
     \figcaption {Trailed {\it IUE} low dispersion spectrum of
        $\alpha$ Aql (SWP 45755, effective exposure time of 3.4 s), and for
  	comparison the UV flux distribution predicted by an ATLAS9 model
	having $T_{\rm eff} = 7900$ K and $\log g = 4$.  The horizontal 
	lines drawn below the spectrum show the wavelength bins for which
	we have measured average fluxes for each star.}
     \end{figure}
     \begin{figure}
     \epsscale{0.8}
     \plotone{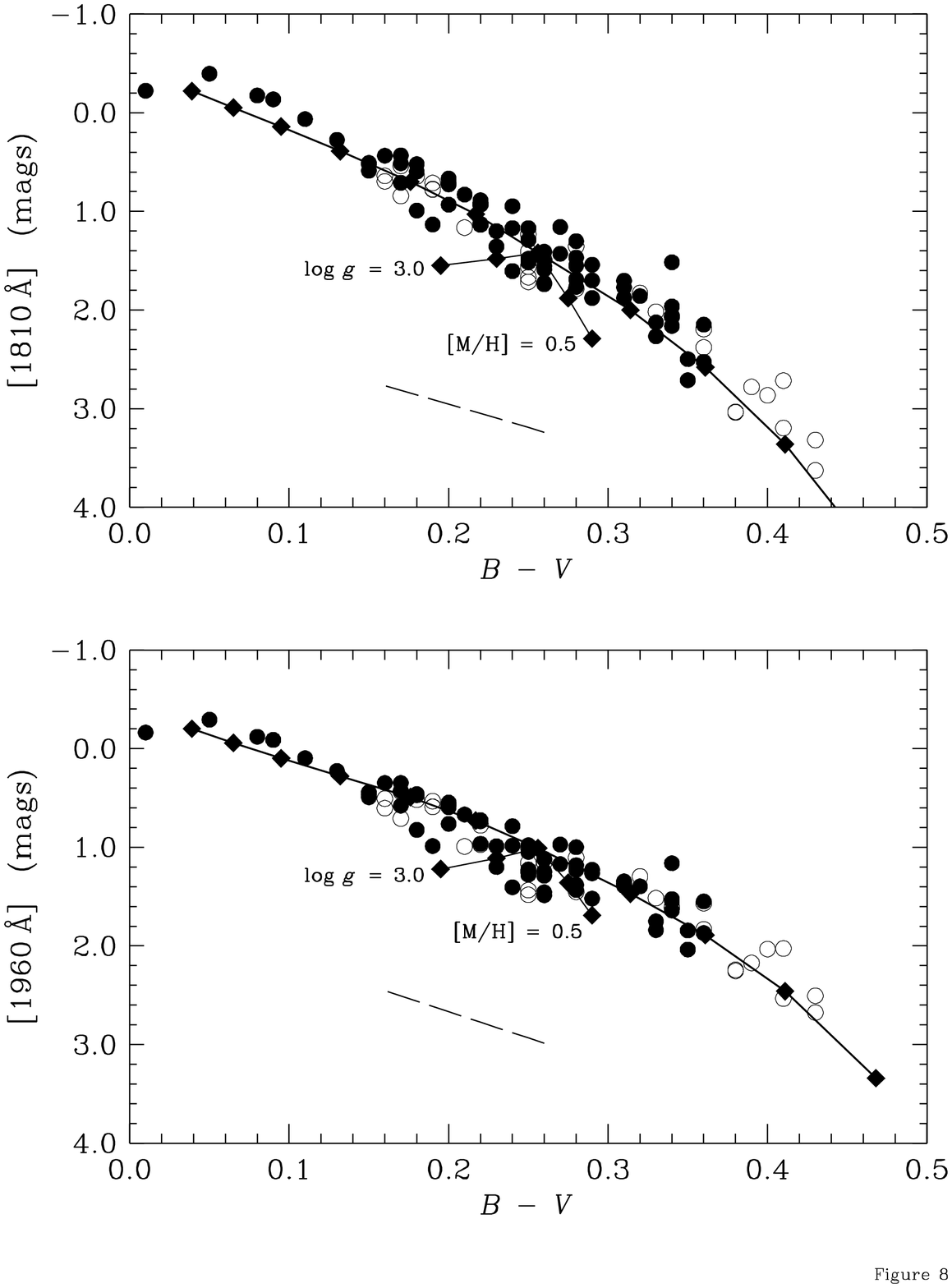}
     \figcaption {Color--color plots for stars observed with {\it IUE}:
	(a) for a wavelength of 1810\AA, (b) for a wavelength of
	1960\AA.  {\it Filled circles}: UV spectra obtained in our
	guest observer program. {\it Open circles}: archival
	data.  The filled diamonds are ATLAS9 models, those extending
	below and to the left of the main sequence illustrating the
	effects of a lower surface gravity (for models with $T_{\rm eff}
	= 7500$ K, solar abundances, and $\log g = 3.5$ and 3.0), and 
	those to the right showing the effects of an increase in the heavy
	element abundances (for [M/H] = 0.3 and 0.5).  The dashed
	line is a normal interstellar reddening line for a color excess
	of E(\bv) = 0.1 mags.}
     \end{figure}
     \begin{figure}
     \plotone{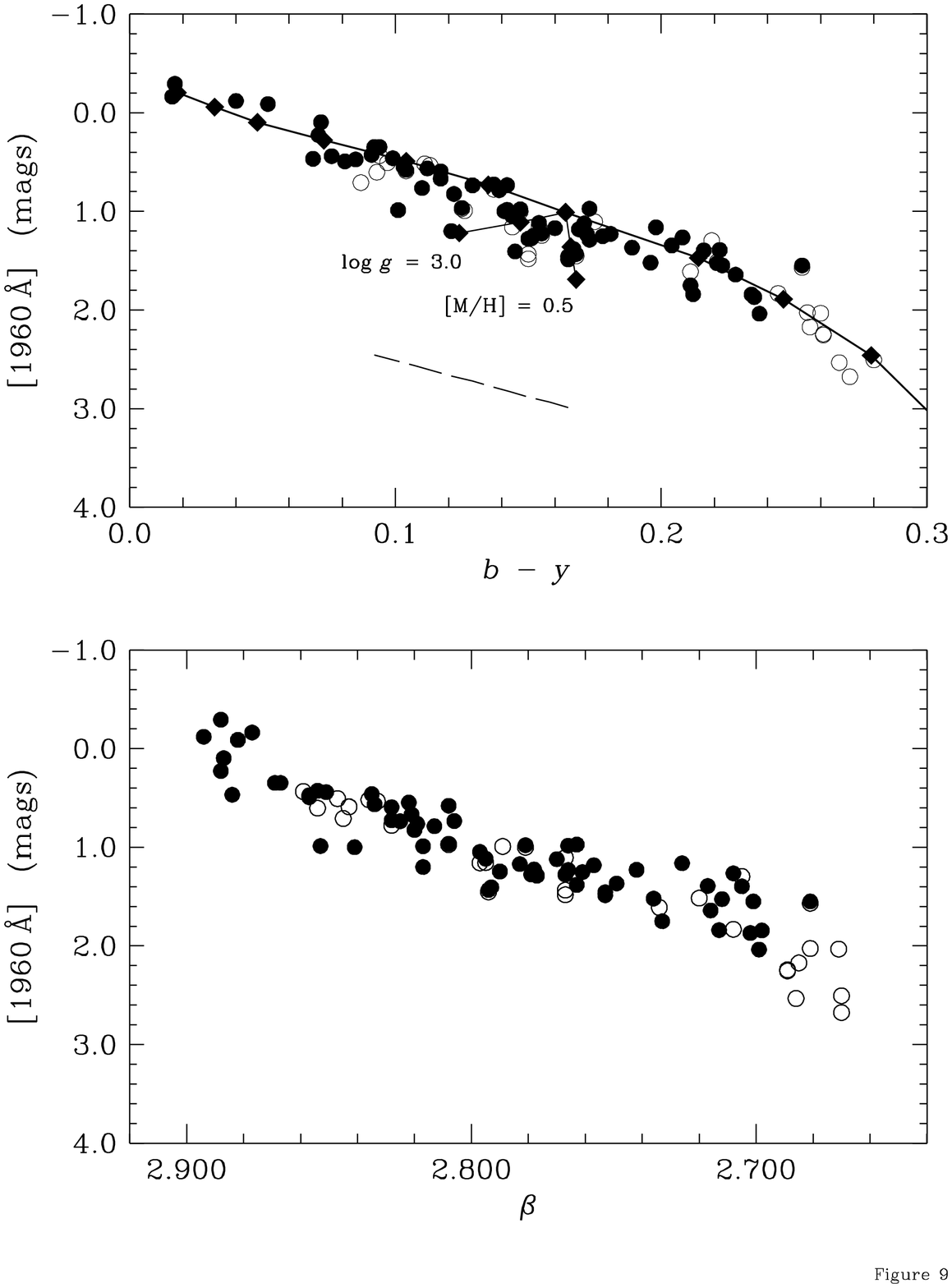}
     \figcaption {Two--color plot with $b - y$ as the abscissa ({\it top
        panel}) and the Str\"omgren H$\beta$ index, $\beta$, as the
        abscissa ({\it bottom panel}).  The symbols are the same as in 
        Fig. 8.  The \bv\ boundaries of the A--star gap correspond 
        approximately to $0.12 \leq b - y \leq 0.21.$ }
     \end{figure}
     \begin{figure}
     \epsscale{0.9}
     \plotone{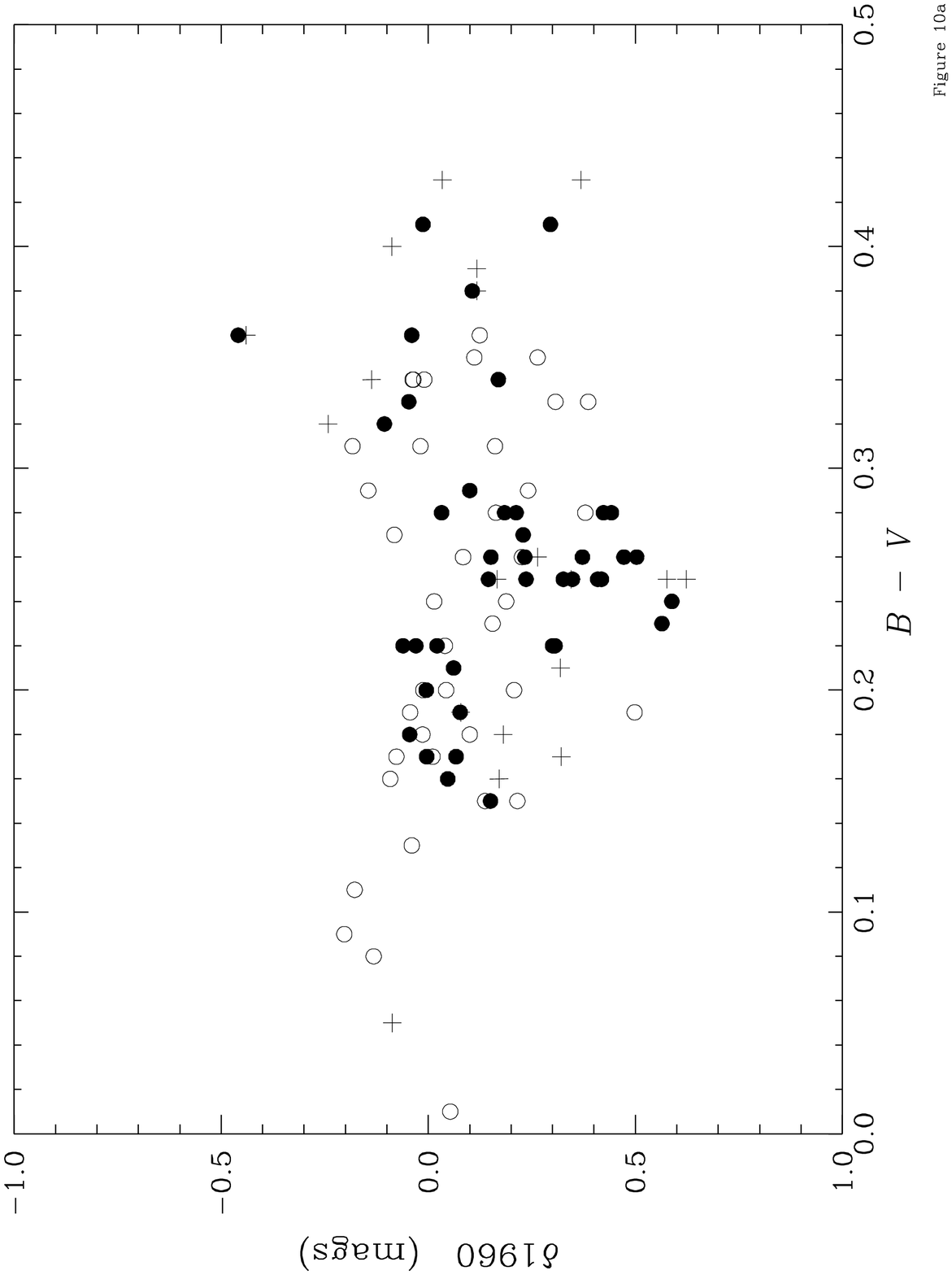}
    \figcaption {(a) The deviations in UV color as a function of imaging
	mode ({\it plus signs}: point source, {\it filled circles}: 
	trailed, {\it open circles}: multiple exposure).  
	(b) Deviations versus $\beta$ angle of the observations 
	({\it filled circles}:  original colors, {\it open circles}:
	luminosity-- and metallicity--corrected colors).  (c) Same as
	(b) for the {\it IUE} SWP camera focus setting.  (d) Same as
	(b) for the maximum DN exposure level in the spectral image. 
	(e) Same as (b) for the exposure time of the image.}
     \end{figure}
     
     \begin{figure}
        \plotone{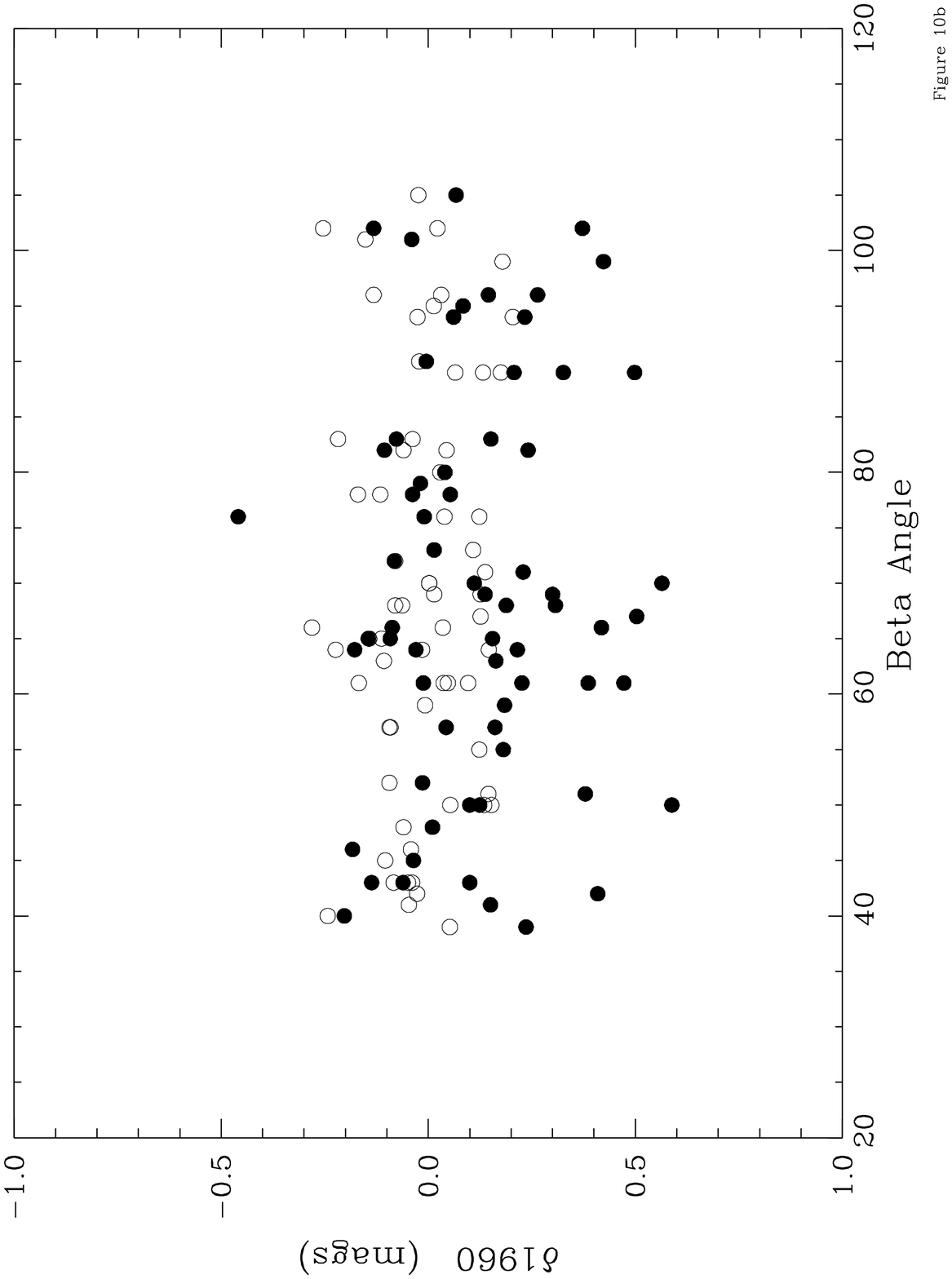}
       \end{figure}
     
      \begin{figure}
      \plotone{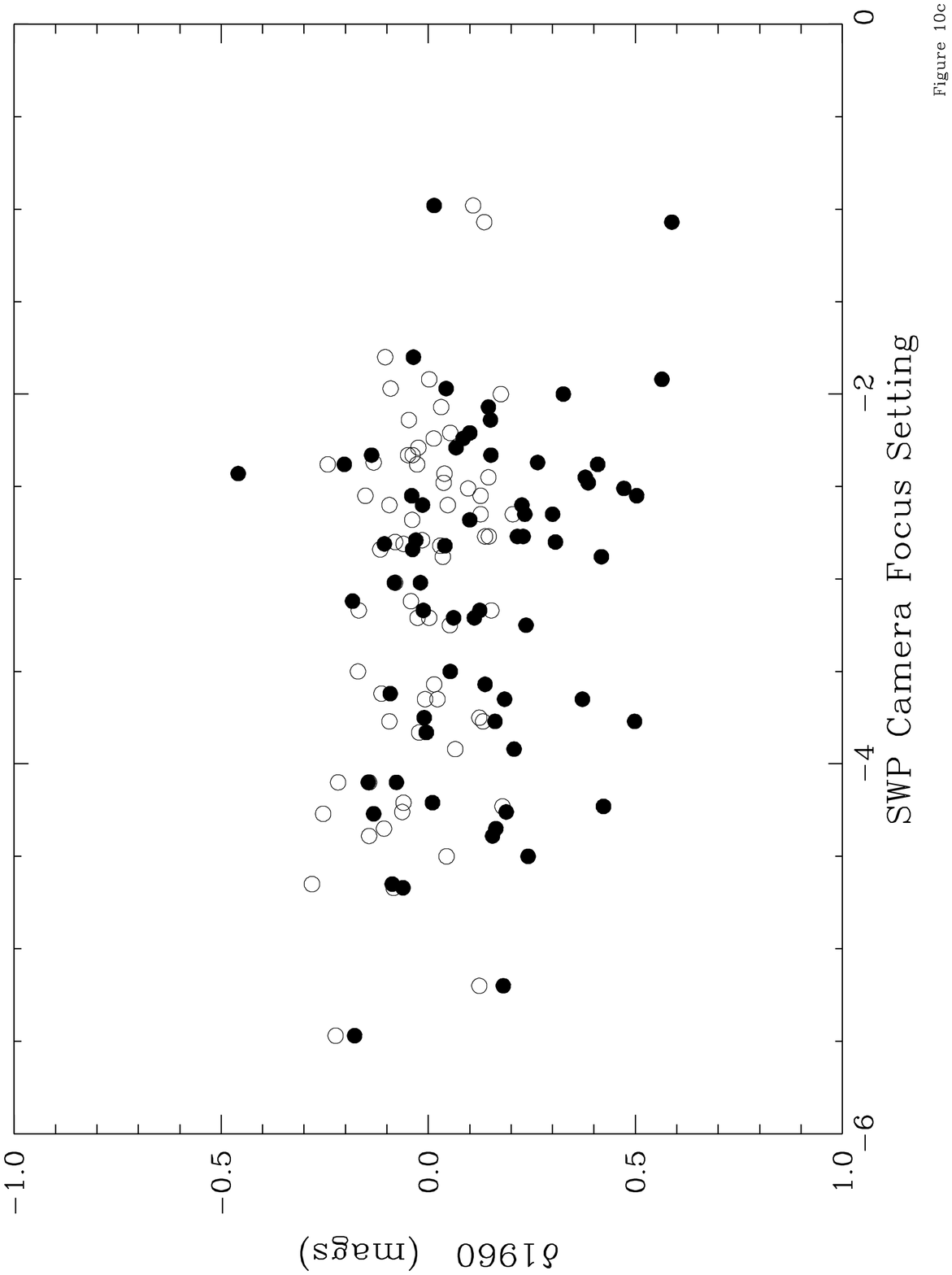}
      \end{figure}
     
      \begin{figure}
     \plotone{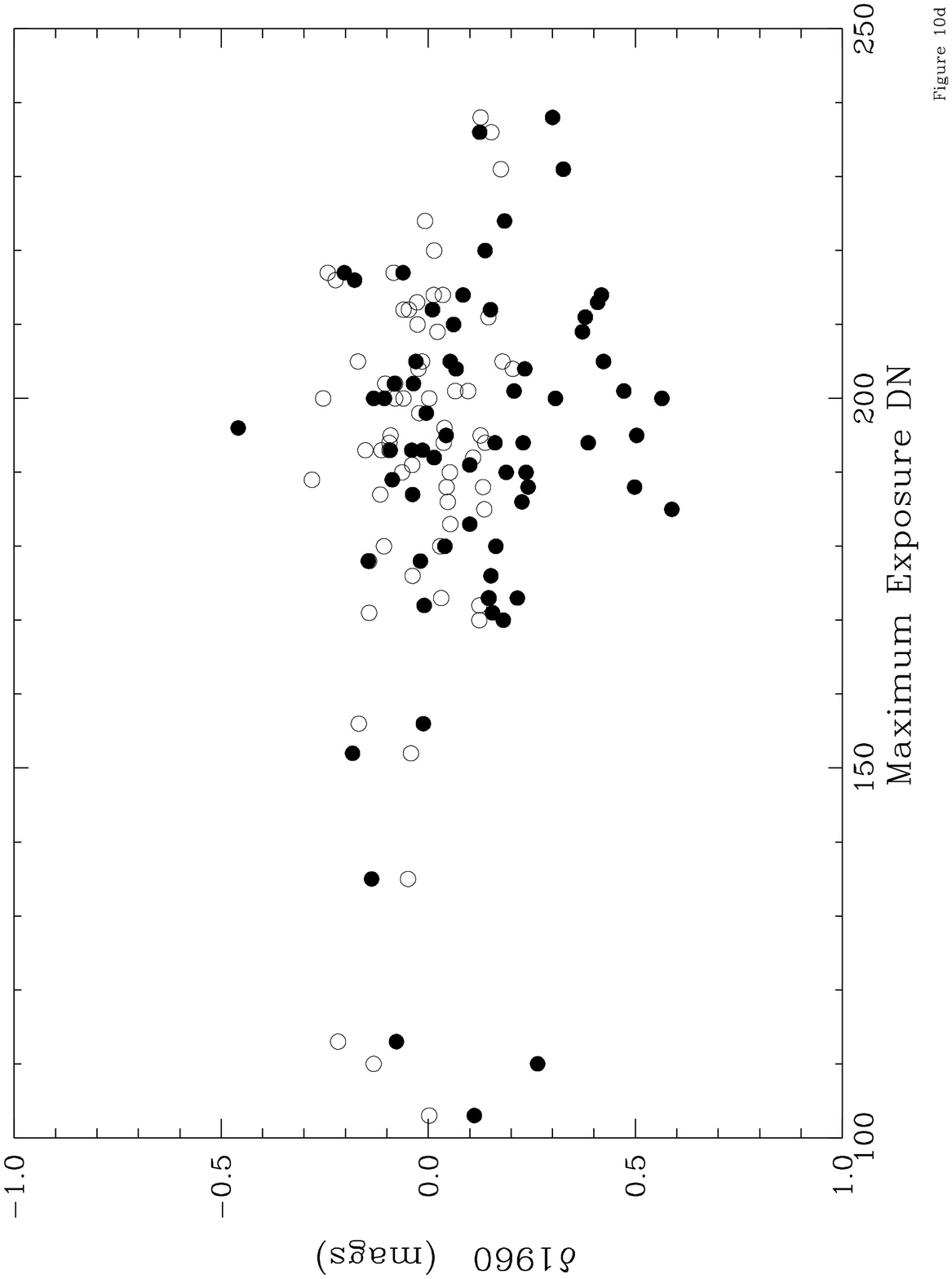}
     \end{figure}
     
     \begin{figure}
     \plotone{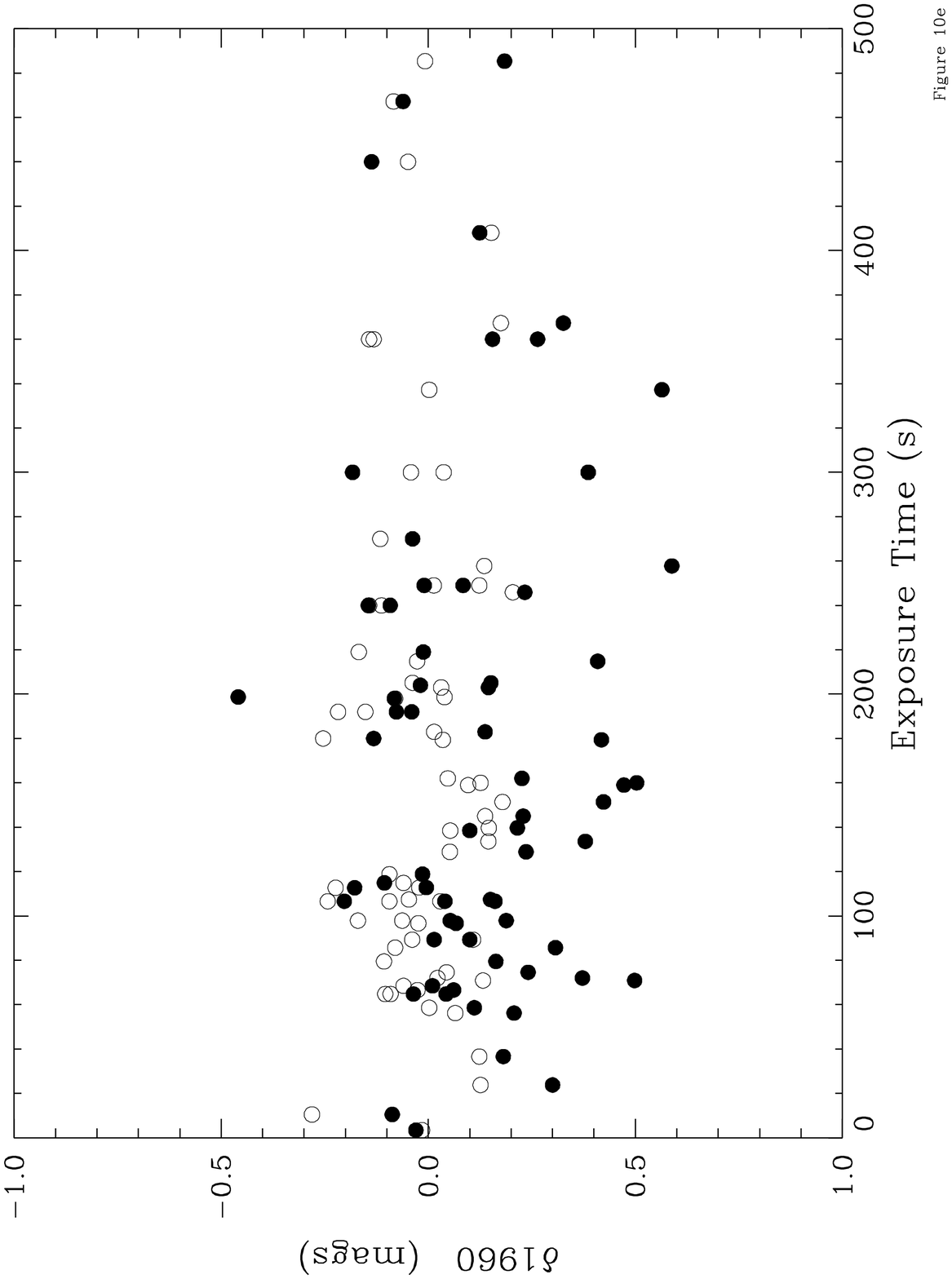}
     \end{figure}
 
     \begin{figure}
     \plotone{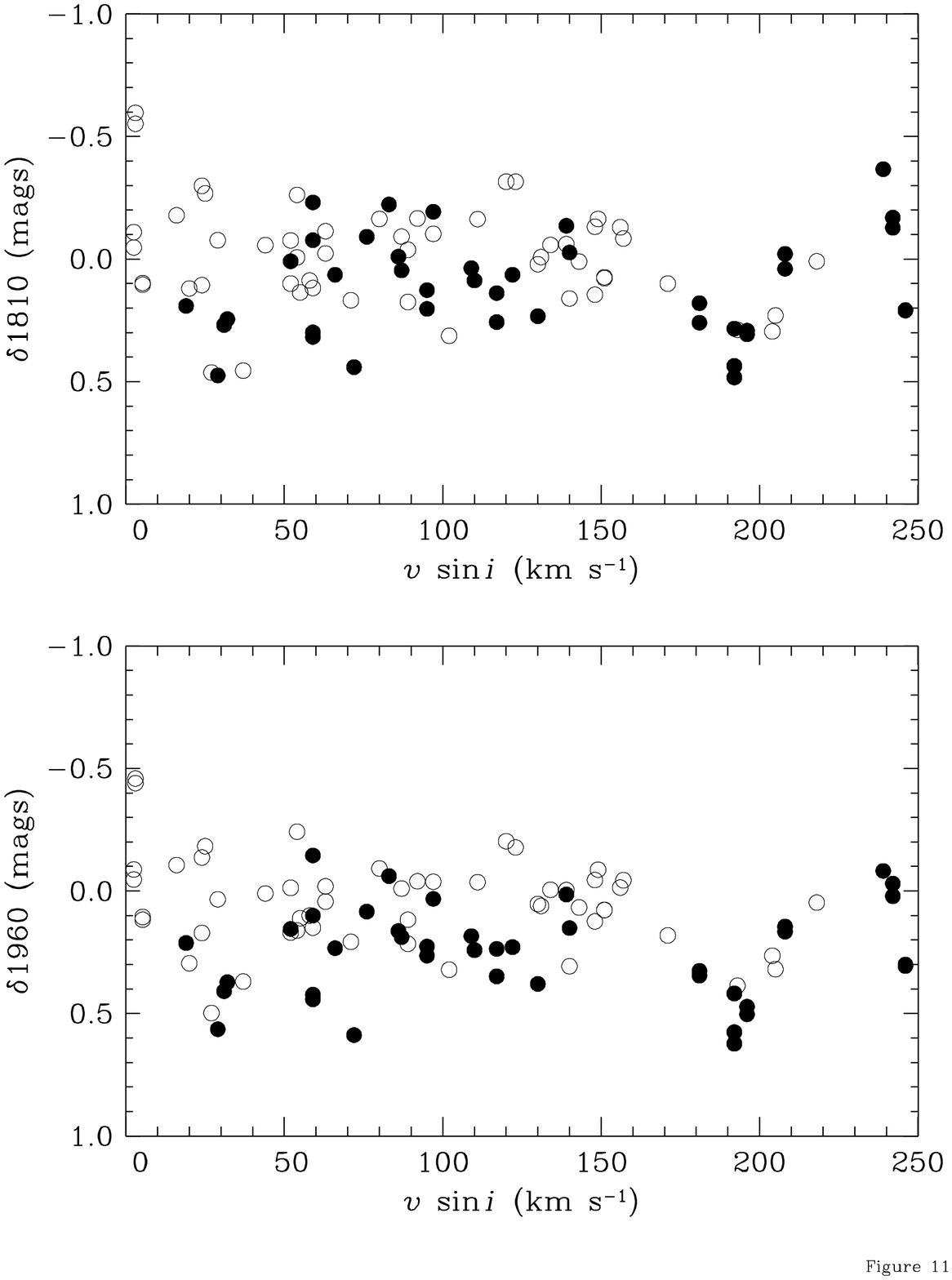}
     \figcaption {(a) The deviations in color at 1810\AA\ versus the stellar
	rotation speed.  {\it Filled circles}:  stars inside the A--star
	gap.  {\it Open circles}:  stars outside the gap.  (b) Same as (a)
	at a wavelength of 1960\AA.}
     \end{figure}
     \begin{figure}
     \epsscale{0.95}
     \plotone{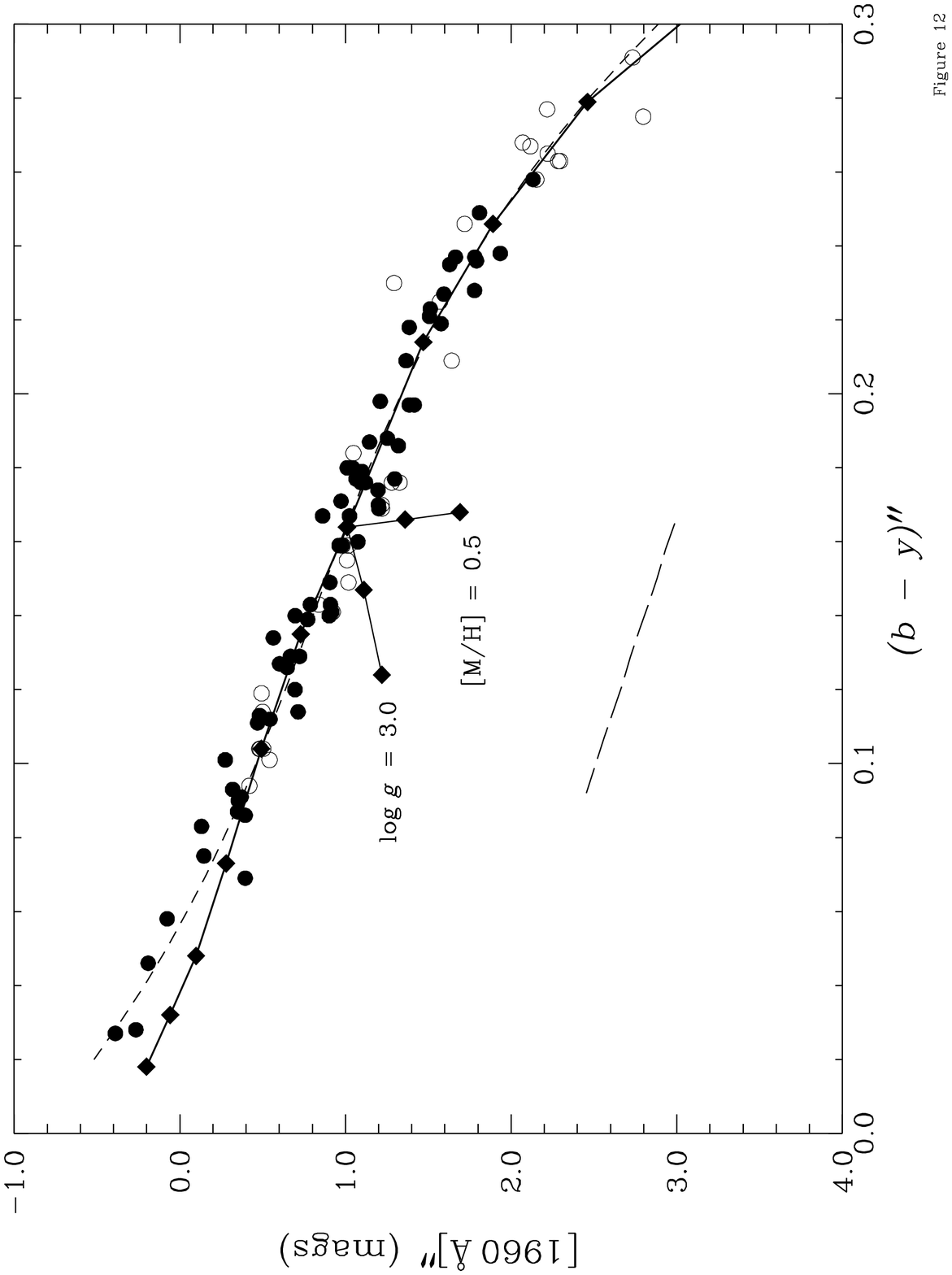}
     \figcaption {Two--color plot for luminosity-- and metallicity--corrected
	colors.  {\it Filled circles}: new spectra obtained here.
	{\it Open circles}: spectra taken from archives.  {\it Filled
	diamonds}:  ATLAS9 models.  The long-dashed line is the interstellar
	reddening line.  The short-dashed line is a cubic polynomial fit to
	the observations.}
     \end{figure}
     \begin{figure} 
     \plotone{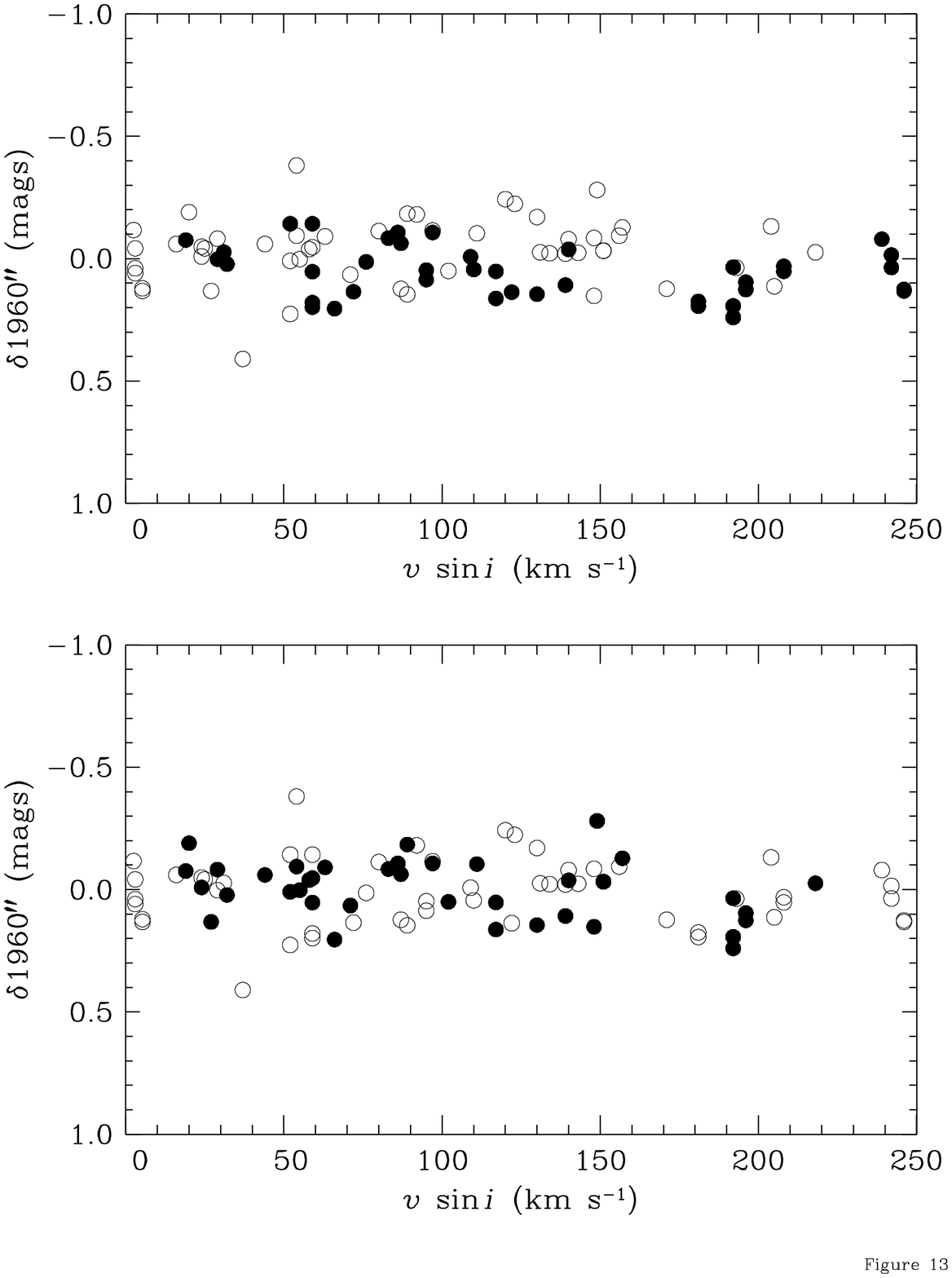}
     \figcaption {Deviations in adjusted 1960\AA\ colors as a function of
	stellar rotation.  (a) for stars inside the A--star gap
        ({\it filled circles}) and those outside the gap ({\it open circles}).
        (b) for double stars ({\it filled circles}) and single stars
        ({\it open circles}).}
     \end{figure}

     \begin{figure}
     \plotone{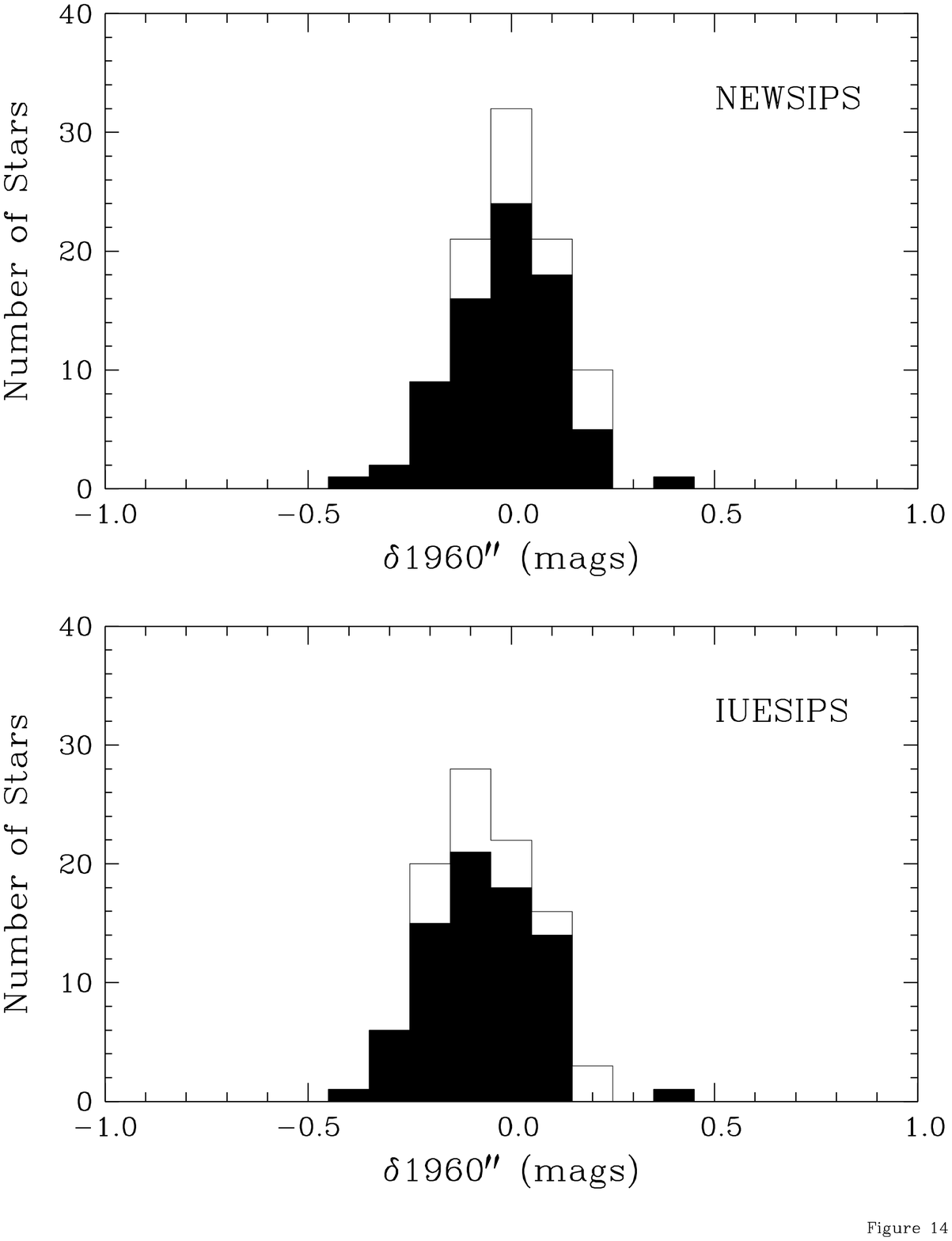}
     \figcaption {A stacked-bar histogram of the deviations in corrected
	1960\AA\ colors for non-variable stars ({\it filled area}) and for 
	known $\delta$ Sct pulsators ({\it open area}).  (a) For the
        NEWSIPS reductions.  (b) For the IUESIPS reductions.}
     \end{figure}

     \end{document}